\documentclass[aps,prb,groupaddress,reprint,twocolomn]{revtex4-1}
\usepackage{amssymb}
\usepackage[pdftex]{graphicx}
\usepackage{amsmath}
\usepackage{amsfonts}
\usepackage{color}
\usepackage{here}
\usepackage{bm}
\usepackage{ulem}

\usepackage{graphicx}
\usepackage{dcolumn}

\begin{document}

  \preprint{APS/123-QED}

  \title{Effect of Element Doping and Substitution on the Electronic Structure and \\
    Macroscopic Magnetic Properties of SmFe$_{12}$-based Compounds}

  \author{Takuya Yoshioka}
  \affiliation{Department of Applied Physics, Tohoku University, Sendai, Miyagi 980-8579, Japan}
  \affiliation{ESICMM, National Institute for Materials Science, Tsukuba, Ibaraki 305-0047, Japan}
  \affiliation{Center for Spintronics Research Network, Tohoku University, Sendai, Miyagi 980-8577, Japan}
  \affiliation{Strategic Technology Center, TIS Inc., Koto, Tokyo 135-0061, Japan}
  \author{Hiroki Tsuchiura}
  \affiliation{Department of Applied Physics, Tohoku University, Sendai, Miyagi 980-8579, Japan}
  \affiliation{ESICMM, National Institute for Materials Science, Tsukuba, Ibaraki 305-0047, Japan}
  \affiliation{Center for Spintronics Research Network, Tohoku University, Sendai, Miyagi 980-8577, Japan}
  \author{Pavel Nov\'ak}
  \affiliation{Institute of Physics of ASCR, Cukrovarnick\'a, Prague 6 162 00, Czech Republic}

  \date{\today}

  \begin{abstract}
    The mechanisms underlying the enhancement of magnetic anisotropies (MAs) of Sm ions, owing to valence electrons at the Sm site and the screened nuclear charges of ligands, are clarified
    using a detailed analysis of crystal fields (CF).
    In order to investigate the finite-temperature magnetic properties,
    we developed an effective spin model for SmFe$_{12}X$ ($X$=H, B, C, and N) and SmFe$_{11}M$ ($M$=Ti, V, and Co),
    where the magnetic moments, CF parameters,
    and exchange fields were determined by first-principle calculations.
    Using this model,
    the MA constants and magnetization curves at finite temperatures
    were investigated using a recently introduced analytical method [T. Yoshioka, H. Tsuchiura, and P. Nov\'ak, Phys. Rev. B {\bf 102}, 184410 (2020)].
    In SmFe$_{12}X$, the doped light elements $X$ are assumed to be at the  $2b$ site,
    and in SmFe$_{11}M$, the substitution site of Fe is systematically investigated for all inequivalent $8f$, $8i$, and $8j$ sites.
    We found that the first-order MA constant $K_1$ is increased by a factor of about two
    when hydrogen is doped to the $2b$ site and when Fe is replaced by Ti or V at the $8j$ site,
    owing to the attraction of the prolate $4f$ electron cloud
    to the screened positive charges of the surrounding ligand ions.
    We found that when Fe is replaced by Co, the MA increases at all temperatures regardless of the substitution site.
    The substituted Co attracts electrons,
    which reduces the electron density in the region from the Sm site to the empty $2b$ site.
    This causes the $4f$ electron cloud at the Sm site to be fixed along the $c$-axis direction, which improves the MA.
    The calculated temperature dependence of $K_1(T)$ and $K_2(T)$ in SmFe$_{11}$Co qualitatively
    reproduces the experimental results in the case of  Sm(Co$_x$Fe$_{1-x}$)$_{12}$ for $x$=0.1 and 0.07.
    The first-order magnetization process is observed at low temperatures in SmFe$_{12}$ itself
    and in many variations of SmFe$_{12}$-based compounds prepared using element doping and substitution.
    This is mainly due to the competition between the conditions $K_1>0$ and $K_2<0$, and that of $K_1<-6K_2$
    owing to the ThMn$_{12}$ structure having a vacancy at the $2b$ site.

  \end{abstract}

  \maketitle


  \section{Introduction}

  Intensive research has been conducted on developing new rare-earth ($R$)
  lean permanent magnetic materials, which have strong magnetic properties comparable to those of Nd-Fe-B.
  Hence, magnetic materials possessing a ThMn$_{12}$ structure,
  which have a high proportion of Fe relative to $R$,
  are again attracting research attention \cite{Miyake,Hirayama1,Hirosawa,Coey,Hadjipanayis}.
  However, the substitution of stabilizing elements is especially important
  \cite{Ohashi,Hu,Kuno,Hirayama2,Schoenhoebel,Ogawa2,Diop,Ogawa,Sepehri-Amin3,Tozman2019,Tang,Makurenkova,Harashima2}.
  For example, nitrogenation of NdFe$_{12}$ compounds improves the magnetic properties considerably
  \cite{Miyake,Harashima2,Hirayama1}.
  Thus, elemental doping and substitution of the ThMn$_{12}$ series of compounds are
  being intensively researched for practical applications,
  and understanding the basic principles governing their behaviour is required, based on the electronic theory.

  \begin{figure}[htb]
    \begin{center}
      \includegraphics[width=8.5cm]{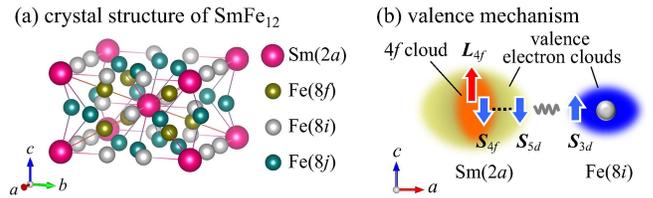} 
      \caption{\label{fig:structval}
        (a) Crystal structure of SmFe$_{12}$ and (b) illustration of the valence mechanism \cite{Tsuchiura1,Yoshioka_SmFe12}.}
    \end{center}
  \end{figure}

  \begin{figure*}[htb]
    \begin{center}
      \includegraphics[width=14cm]{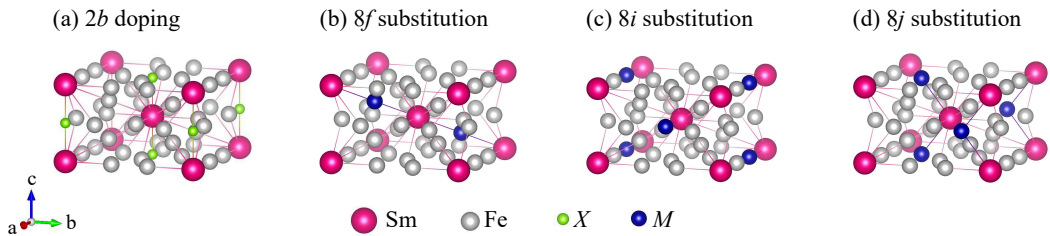} 
      \caption{\label{fig:struct}
        Crystal structures of (a) SmFe$_{12}X$ ($X$=H, B, C, and N) for $2b$-site doping and
        SmFe$_{11}M$ ($M$=Ti, V and Co) with substitution for each Fe (b) $8f$, (c) $8i$, and (d) $8j$ sites in SmFe$_{12}$.
      }
    \end{center}
  \end{figure*}

  Recently, a thin film of SmFe$_{12}$ was synthesized and
  its intrinsic magnetic properties were investigated experimentally \cite{Hirayama2}.
  The magnetic properties of SmFe$_{12}$ can be improved
  by substituting Co for Fe \cite{Hirayama2,Ogawa,Ogawa2} and the addition of B increases its coercivity,
  which has attracted much attention in terms of its applicability \cite{Sepehri-Amin3}.
    On the other hand, theoretical studies have long been conducted \cite{KuzminDy,Harashima1,Harashima2,Koerner,Ke,Delange,Yoshioka_SmFe12}.
Using a simple model, Kuz'min $et$ $al$. \cite{KuzminDy} found that the leading term of the CF parameter $A_2^0$ is negative for Fe-based $R$Fe$_{12-x}M_x$ compounds regardless of the type of $R$.
Harashima et al \cite{Harashima1,Harashima2}. and K\"orner $et$ $al$. \cite{Koerner} investigated the effect of nitridation on the magnetic properties of a series of 1-12.
For SmFe$_{12}$, it was found that nitridation changes the sign of $A_2^0$ from negative to positive.
Ke $et$ $al$. \cite{Ke} have extensively investigated the effects of doping and substitution systematically for $R$Fe$_{12}$-based compounds for $R$=Y and Ce.
Their study revealed that the Fe dominated transition metal sublattice has uniaxial magnetic anisotropy  in the range of the investigated compounds, which are YFe$_{12}$ and YFe$_{11}$Ti$X$ ($X$=H, C, N).
The above analyses are based on the simple model or first-principles calculations at absolute zero.
Delange $et$ $al$. \cite{Delange} studied the effect of N and Li doping on the finite temperature MA of SmFe$_{12}$ in detail.
However, the effect of substitution, which is important in the 1-12 system, was not taken into account
and magnetization curves that can be compared to experiments were not presented.
We calculated the MA constants at finite temperatures as well as the magnetization curves for the 1-12 system \cite{Yoshioka_SmFe12}.
However, we did not discuss the effect of doping and substitution on the SmFe$_{12}$.
Therefore, detailed analysis of the finite-temperature magnetic properties must be performed, 
including magnetic anisotropy constants and magnetization curves of SmFe$_{12}$ compounds with doping and substitution.
  In this study, we analyze the electronic states of SmFe$_{12}$-based compounds
  by considering element doping and substitution independently.
  In addition, we construct an effective spin model based on
  theses electronic states and analyze their macroscopic magnetic properties at finite temperatures.

  FIG. \ref{fig:structval} shows (a) the crystal structure and (b) the mechanism underlying MA.
  In our previous paper \cite{Yoshioka_SmFe12},
  we reported that the uniaxial MA in SmFe$_{12}$ is a consequence of the Coulombic interaction
  between the valence and $4f$ electron clouds.
  However, a quantitative and comprehensive understanding, based on the electronic structure that has undergone doping and substitution,
  of the mechanism underlying the MA is still insufficient, which is essential for the efficient application of SmFe$_{12}$-based compounds.

  The purpose of this study was to extract the macroscopic magnetic properties of SmFe$_{12}$-based compounds
  at finite temperatures based on electronic theory using a recently developed reliable method
  \cite{Yoshioka_SmFe12,Tsuchiura1,Yoshioka,Tsuchiura2,Yamashita}.
  In this study, we focus on the SmFe$_{12}X$ ($X$=H, B, C, and N) and SmFe$_{11}M$ ($M$=Ti, V, and Co) compounds.
  The electron structure and finite-temperature magnetic properties are
  investigated within the linear theory for CF Hamiltonian
  \cite{Yoshioka_SmFe12}.

  This paper is organized as follows.
  In {\S II}, we introduce the effective spin model
  and analytical method for SmFe$_{12}$-based compounds.
  In {\S III}, we briefly outline the mechanism underlying the MA on $R$ ion in $R$-transition-metal compounds.
  In {\S VI}, we present the doping and substitution effects on the electronic structure and MA.
  The macroscopic MA constants at finite temperatures and magnetization curves, along with a brief summary are detailed in §V.

  \section{Model and Method}
  In this study,
  we investigate the macroscopic magnetic properties from the electronic structure
  of the SmFe$_{12}X$ ($X$=H, B, C, and N) and SmFe$_{11}M$ ($M$=Ti, V, and Co) shown in FIG. \ref{fig:struct}.

  For this purpose, we determine the model parameters using first-principle calculations and construct an effective spin model.
  In this section, we describe the model Hamiltonian and the method to determine its parameters.
  Next, we utilize the method to investigate the magnetic properties at finite temperatures.

  \subsection{Model Hamiltonian for $R$-Transition-Metal Compounds}
  In order to investigate the bulk magnetic properties,
  we assume homogeneity of the partial magnetization $M^{\rm TM}$ and
  partial anisotropy constant $K_1^{\rm TM}$, excluding the contribution of the $4f$ electrons.
  The parameters $M^{\rm TM}$ and $K_1^{\rm TM}$ are mainly associated with the transition-metal elements.
  To consider the effect of a small amount of elemental doping and substitution with respect to SmFe$_{12}$,
    we assume a similar uniaxial MA of SmFe$_{11}$Ti, which has been confirmed experimentally \cite{Hu,Nikitin}.
  In this case, the Hamiltonian of the system can be written in terms of
  the following two-sublattice model \cite{Yoshioka_SmFe12,Yoshioka,Hummler0,Yamada}:
  \begin{align}
    \hat{\mathcal{H}}=&\sum^{n_R}_{j=1}\hat{\mathcal{H}}_{R,j}\nonumber\\
    &+VK_{1}^{\rm TM}(T)
        \left[1-\left({\bm M}^{\rm TM}\cdot {\bm n}_c\right)^2\right]
        -V{\bm M}^{\rm TM}(T)\cdot\boldsymbol{B}, \label{eq:Htot}
  \end{align}
  where $\hat{\mathcal{H}}_{R,j}$ is a Hamiltonian for single $R$ ion
  at the site $j$, ${\bm M}^{\rm TM}(T)$ is the partial
  magnetization excluding the contribution of the $R$ ions,
  ${\bm n}_c$ is the unit vector along the $c$-axis,
  and $n_R$ is the number of $R$ ions in the volume of the cell $V$ to be considered.
  Hereafter $j$ is omitted for simplicity.
  The $\hat{\cal H}_{R}$ can be written as \cite{Yoshioka_SmFe12,Sankar,Wijn}:
  \begin{align}
    \hat{\cal H}_{R}=&\hat{\cal H}_{{\rm so}}+\hat{\cal H}_{{\rm ex}}+\hat{\cal H}_{{\rm CF}}+\hat{\cal H}_{{\rm Z}},\label{eq:Heff}\\
    &\hat{\cal H}_{{\rm so}}=\xi\sum_{i=1}^{n_{4f}} \hat{\bm l}_i\cdot\hat{\bm s}_i,\\
    &\hat{\cal H}_{{\rm ex}}=2\mu_B{\bm B}_{{\rm ex}}(T)\cdot\sum_{i=1}^{n_{4f}}\hat{\bm s}_i,\label{eq:Bex}\\
    &\hat{\cal H}_{{\rm CF}}=\sum_{l,m}\frac{A_{l}^{m}\langle r^{l}\rangle}{a_{l,m}}\sum_{i=1}^{n_{4f}}
    t_{l}^{m}(\hat{\theta}_{i},\hat{\phi}_{i}),\\
    &\hat{\cal H}_{{\rm Z}}=\mu_B{\bm B}\cdot\sum_{i=1}^{n_{4f}}(\hat{\bm l}_i+2\hat{\bm s}_i),
  \end{align}
  where the summation of $\sum_{i=1}^{n_{4f}}$ is taken over $4f$ electrons in the $R$ ion.
  $\hat{\cal H}_{{\rm so}}$ is the spin-orbit interaction
  between the spin ($\hat{\bm s}_i$) and orbital ($\hat{{\bm l}_i}$) angular momenta, with a coupling constant $\xi$.
  $\hat{\cal H}_{{\rm ex}}$ is the exchange interaction between the spin moment and temperature-dependent exchange field
  ${\bm B}_{{\rm ex}}(T)$, where $\mu_B$ is the Bohr magneton.
  $\hat{\cal H}_{{\rm CF}}$ is the CF Hamiltonian, where
  $A_{l}^m\langle r^l\rangle$ is the CF parameter for the $j$-th $R$ site,
  $a_{l,m}$ is a numerical factor \cite{Stevens,Hutchings},
  $t_l^m(\hat{\theta}_i,\hat{\varphi}_i)$ is the
  tesseral harmonic function of the polar and azimuthal angle $\hat{\theta}_i$ and $\hat{\varphi}_i$.
  $\hat{\cal H}_{{\rm Z}}$ is the Zeeman term with an applied field ${\bm B}$.

  The hierarchy of the energy scale of each term is as follows:
  \begin{equation}
    \hat{\cal H}_{\rm so}\gg\hat{\cal H}_{\rm ex}\gg\hat{\cal H}_{\rm CF}\sim\hat{\cal H}_{\rm Z}.
  \end{equation}
  In this situation, it is possible to map the density functional theory (DFT) to the CF theory within the first-order of the CF term \cite{Faehnle1995}.
  Therefore, we apply our linear theory for the CF \cite{Yoshioka_SmFe12}
  to the two-sublattice model to clarify the finite temperature magnetic properties,
  where we apply the $LS$ coupling scheme with the assumption of a trivalent $R$.
  In the treatment of $\hat{\cal H}_{\rm so}$, we should note that
  because the $LS$ coupling in Sm compounds is weak compared with those of the other $R$,
  the excited $J$-multiplets must be included\cite{Yoshioka_SmFe12,VanVleck,Sankar,Wijn,Kuzmin_mix,Magnani}.
  Details are shown in Appendix \ref{Sec:HR}.


  For SmFe$_{12}X$ ($X$=H, B, C, and N),
    the doped light elements are assumed to be at the $2b$ site [FIG. \ref{fig:struct} (a)].
  In this case, the lattice constants as for SmFe$_{12}$ \cite{Hirayama2} that is,
  $a=b=$8.35 \AA\ and $c$=4.8 \AA\ are used.
  The substituted compounds of SmFe$_{11}M$ ($M$=Ti, V, and Co) are systematically
  investigated for all possible replacement sites $8f$, $8i$, and $8j$ for each $M$.
  The experimentally determined lattice constants are used for the first-principle calculations,
  where $a=b=$8.54 \AA\ and $c$=4.78 \AA\ \cite{Hu}
  , $a=b=$8.5205 \AA\ and $c$=4.7693 \AA\ \cite{Schoenhoebel}
  , and $a=b=$8.4 \AA\ and $c$=4.8 \AA\ \cite{Hirayama2}, for $M$=Ti, V, and Co, respectively.

  Local distortions due to doping and substitution are expected to affect the results.
  In this study, however, we focus on the effect of chemical changes on the MA;
  thus, we simplified the internal structure by fixing it to the SmFe$_{12}$ as
  (0.25,0.25,0.25) for $8f$, (0.359,0.000,0.000) for $8i$, and (0.270,0.500,0.000) for $8j$ sites \cite{Harashima1}.
  For reference, the energies of the systems for SmFe$_{11}M$ studied in this study are shown
  in Appendix \ref{Sec:energy}.
  The results are comparable to those of the previous theoretical study \cite{Harashima3}.

  \subsection{Model Parameters Determined by First-Principle Calculations}

  In this study, the model parameters are determined using DFT calculations.
  We use the full-potential linearized augmented plane wave plus local orbitals (APW+lo)
  method implemented in the WIEN2k code \cite{wien2k}. The Kohn-Sham equations
  are solved within the spin polarized generalized-gradient approximation (SGGA).
  The $4f$ electron states cannot be correctly described by
  local or semi-local approximations to DFT.
  Therefore, we treat them as core states of an atom, which corresponds to the so-called opencore method
  \cite{Novak,Richter1,Richter2,Hummler2,Divis1,Divis2}.

  The total energy of the system in the DFT,
  as a function of the total charge density $\rho({\bm r})$, has the following form\cite{Hohenberg,Faehnle1995}:
  \begin{equation}
    E[\rho({\bm r})]=T_s[\rho({\bm r})]+E_{\rm en}[\rho({\bm r})]+E_{\rm H}[\rho({\bm r})]+E_{\rm XC}[\rho({\bm r})],
  \end{equation}
  Where the right hand side represents a sum of the non-interacting kinetic,
  electron-nucleus, Hartree, exchange-correlation (XC) energies.
  $\rho({\bm r})$ is decomposed to $\rho({\bm r})=\rho_{4f}(\bm r)+\rho_{\rm rest}(\bm r)$,
  where $\rho_{\rm rest}(\bm r)$ is the total electron density excluding the $4f$ electrons.

  The change in the CF energy
  within the atomic sphere radius $r^R_{\rm AS}$ with respect to the change
  in the orientation of the $4f$ electron cloud $\Delta \rho_{4f}$ can be written as \cite{Faehnle1995,Kohn}
  \begin{align}
    \Delta E_{\rm CF}
    &=\int_{|{\bm r}|<r^{R}_{\rm AS}}
    \left\{V_{\rm en}({\bm r})+V_{\rm H}[\rho_{\rm rest}^0({\bm r})]\right\}
    \Delta \rho_{4f}({\bm r})d^3r,\label{DHCF1}
  \end{align}
  where $V_{\rm en}(\bm r)$ and $V_{\rm H}[\rho_{\rm rest}^0(\bm r)]$ are
  the Coulomb potential of the nuclei and the Hartree potential, respectively,
  which can be written in the following form \cite{Kohn}:
  \begin{align}
    &V_{\rm en}({\bm r})=\frac{\delta E_{\rm en}[\rho({\bm r})]}{\delta \rho(\bm r)}
    =-\frac{e^2}{4\pi\varepsilon_0}\sum_n\frac{Z_n}{|{\bm r}-{\bm R}_n|},\\
    &V_{\rm H}[\rho_{\rm rest}^0(\bm r)]=
    \left.\frac{\delta E_{\rm H}[\rho({\bm r})]}{\delta \rho(\bm r)}\right |_{\rho=\rho^0_{\rm rest}}
    =\frac{e^2}{4\pi\varepsilon_0 }
    \int\frac{\rho^0_{\rm rest}({\bm r}')}{|{\bm r}-{\bm r}'|}d^3r',
  \end{align}
  where $Z_n$ is the nuclear charge on the $n$-th site and $\rho_{\rm rest}^0({\bm r})$ is the total charge
  excluding the $4f$ electrons, which is fixed to the self-consist charge density.
  Moreover, the change in the expectation value of the CF Hamiltonian can be written as:
  \begin{align}
    \Delta \langle \hat{\cal H}_{\rm CF}\rangle_{4f}
    =&\sum_{l,m}\frac{A_{l}^{m}\langle r^{l}\rangle}{a_{l,m}}
    \Delta\left\langle\sum_{i=1}^{n_{4f}}
    t_{l}^{m}(\hat{\theta}_{i},\hat{\phi}_{i})\right\rangle_{4f}
    \label{DHCF2}
  \end{align}
  By comparing Eqs. (\ref{DHCF1}) and (\ref{DHCF2}),
  the CF parameters can be obtained in the following form:
  \begin{align}
    A_{l}^{m}\langle r^{l}\rangle =&a_{l,m}\int_{|{\bm r}|<r^R_{\rm AS}}d^3r
    |R_{4f}(r)|^2t_l^m(\theta,\phi)\nonumber\\
    &\times \left\{V_{\rm en}({\bm r})+V_{\rm H}[\rho_{\rm rest}^0({\bm r})]\right\}
    \label{eq:Alm}
  \end{align}
  where $R_{4f}(r)$ is the radial part of the $4f$ wave function
  and $r^R_{\rm AS}$ denotes the radius of an atomic sphere.
  As shown in Eq. (\ref{eq:Alm}), the aspherical part of $V_H[\rho_{\rm rest}^0({\bm r})]$
  denotes the CF acting on the $4f$ electrons.
  In the actual calculation,
  the opencore method is applied to treat the $4f$ electrons as spherical core electrons.
  In this case, $\rho_{\rm rest}^0({\bm r})$ in Eq. (\ref{eq:Alm})
  can be replaced with the total charge density $\rho^0({\bm r})$, obtained using self-consistent calculations.

  The exchange field $B_{\rm ex}(T)$ at $T=0$ K acting on the $4f$ spin moments
  in the two-sublattice model in Eq. (\ref{eq:Bex}) can be obtained by comparing
  the change in the total energy $\Delta E$ in DFT calculations with
  one in the expectation value of the exchange term $\hat{\cal H}_{\rm ex}$:
  \begin{align}
    \Delta\langle\hat{\mathcal{H}}_{\rm ex}\rangle_{4f}  &  =
    2\mu_{\rm B}{\bm B}_{\rm ex}(0)\cdot
    \Delta\left\langle\sum_{i=1}^{n_{4f,j}}\hat{\bm s}_i\right\rangle_{4f},
    \label{Hex}
  \end{align}
  where $\mu_{\rm B}$ is the Bohr magneton.
  Since, the opencore method allows us to control the number of occupied $4f$ electrons for each spin,
  it is possible to estimate the increase in energy when the total spin is rotated by 180°,
  within Hund's first rule.
  If the increase of total energy due to the spin flip is denoted as $\Delta E$,
  $B_{\rm ex}$ can be obtained from the following equation\cite{Brooks,Liebs1,Liebs2}:
  \begin{equation}
    B_{\rm ex}(0)=\Delta E/4\mu_{\rm B}S,
  \end{equation}
  where $S$ is the total spin angular momentum of the $4f$ electrons.
  The partial magnetization $M^{\rm TM}(T)$ at $T=0$ K in Eq. (\ref{eq:Htot})
  is obtained in the framework of SGGA as:
  \begin{equation}
    VM^{\rm TM}(0)=
    \int_{V}\left[\rho_{{\rm rest},\uparrow}^0({\bm r})-\rho_{{\rm rest},\downarrow}^0(\bm r)\right]dr^3,
    \label{eq:MTM}
  \end{equation}
  where $\rho_{{\rm rest},\sigma}^0$ is the self-consistent charge density for spin $\sigma$ excluding the $4f$ electrons.

   The following are the technical parameters required for electronic structure calculations using WIEN2k code (version 16.1).
    The number of $k$ points within the whole Brillouin zone was set to 8$\times$8$\times$14.
    The momentum space integrations are performed using the linear tetrahedron method with Bl\"ochl correction.
    For SmFe$_{12}$H, 1406 basis functions with RKmax=4.0 were used, and for all the other compounds 6063 basis functions with RKmax=6.0 were used.
    The radius for Sm ions is $r^{\rm Sm}_{\rm AS}=3.2 a_0$,
    for $M$=Fe, Ti, V, and Co ions is $r^M_{\rm AS}=2.21 a_0$,
    and for $X$=H, N, B, and C ions is $r^X_{\rm AS}=1.32 a_0$,
    where $a_0$ is the Bohr radius.

    In analyzing the MA,
    it is important to investigate the dependence of the CF parameters on the $r^{\rm Sm}_{\rm AS}$ for the series of systems.
    However, due to the significant computational resources required for determining this dependence,
    the maximum possible radius for Sm
    within which the atomic radii do not overlap,
    $r^{\rm Sm}_{\rm AS}=3.2a_0$, was adopted in this study. 
  
  \subsection{Phenomenological Parameters}

  For temperature dependence of $M^{\rm TM}(T)$ and $K_1^{\rm TM}(T)$,
  we apply the phenomenological
  Kuz'min formula \cite{Kuzmin_mag} and extended power low \cite{Miura_pl}
  under the uniformity assumption as:
  \begin{align}
    \frac{M^{\rm TM}(T)}{M^{\rm TM}(0)}  &  =\alpha(T),\label{eq:Kuzmin}\\
    \frac{K_{1}^{\rm TM}(T)}{K_{1}^{\rm TM}(0)}&=\alpha^{3}(T)+\frac{8}{7}C_{1}\left[
      \alpha^{3}(T)-\alpha^{10}(T)\right]  \nonumber\\&+\frac{8}{7}C_{2}\left[  \alpha
      (T)^{3}-\frac{18}{11}\alpha(T)^{10}+\frac{7}{11}\alpha(T)^{21}\right],\label{eq:K1TM}\\
    \alpha(T)  &  =\left[  1-s\left(  \frac{T}{T_{\mathrm{C}}}\right)
      ^{3/2}-(1-s)\left(  \frac{T}{T_{\mathrm{C}}}\right)  ^{5/2}\right]  ^{1/3}\label{eq:alpha},
  \end{align}

  As shown above, $M^{\rm TM}(0)$ can be obtained
  from the first-principle calculations according to the Eq. (\ref{eq:MTM}),
  and for the MA constant $K_1^{\rm TM}(0)$,
  we use the experimental value of YFe$_{11}$Ti at low temperatures as given in Ref. \cite{Nikitin}.
  $T_{\rm C}$ and $s$ are the fitting parameters which are used
  to reproduce the experimental results of the temperature-dependent spontaneous magnetization.

\begin{figure}[H]
    \begin{center}
      \includegraphics[width=8.0cm]{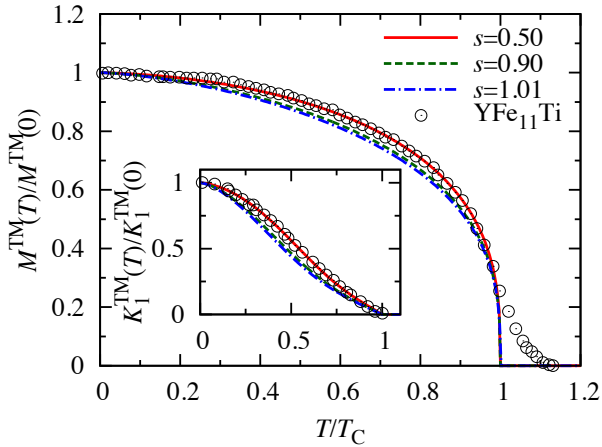}
      \caption{\label{YFe11Ti}
        TM-sublattice magnetization $M^{\rm TM}(T)$ and MA constant $K_1^{\rm TM}(T)$
        (inset) for $s=0.50$, $0.90$ \cite{Hirayama2}, and $1.01$ \cite{Hirayama2},
        used in the present calculations,
        where $s$ is the shape parameter of the Kuz'min formula \cite{Kuzmin_mag}.
        Open circles are experimental results of $M^{\rm TM}(T)$ and $K_1^{\rm TM}(T)$
        in YFe$_{11}$Ti compounds \cite{Nikitin},
        both of which are well fitted by $s=0.5$, $C_1=-0.263$, and $C_2=-0.237$
        in Eq. (\ref{eq:Kuzmin}) and (\ref{eq:K1TM}).}
    \end{center}
  \end{figure}

  For SmFe$_{12}$, the values of $T_{\rm C}$=555 K and $s$=1.01
  are used, which were determined by Hirayama et al. \cite{Hirayama2}.
  For SmFe$_{12}X$ ($X$=H, B, C, and N),
  we use the dimensionless parameter $T/T_{\rm C}$ and $s$=1.01,
  because experimental values of $T_{\rm C}$ do not exist.
  For SmFe$_{11}M$ with an elemental substitution of $M$=Ti, V, and Co,
  we use $T_{\rm C}$=584\cite{Hu}, 634\cite{Schoenhoebel}, and 710\cite{Hirayama2} K as the experimental value of Curie temperatures and
  $s$=0.5, 0.5, and 0.9\cite{Hirayama2} as the form factors in Eq. (\ref{eq:alpha}),
  respectively, where $s$=0.5 used for SmFe$_{11}$Ti and SmFe$_{11}$V
  is determined from YFe$_{11}$Ti as shown in FIG. \ref{YFe11Ti}.
  We note here that the experimental value of $T_{\rm C}$ and $s$
  for SmFe$_{10.8}$Co$_{1.2}$ in Ref. Hiarayama $et$ $al$. \cite{Hirayama2}
  have been adopted as the model parameters for SmFe$_{11}$Co.

The MA constant $K_1^{\rm TM}(T)$ in Eq. (\ref{eq:K1TM}) is expressed by the parameters $s$, $T_{\rm C}$, $K_1^{\rm TM}(0)$, $C_1$, and $C_2$.
  Among these, the parameters $s$ and $T_{\rm C}$ acquire material-specific values given in the previous paragraph.
  For remaining parameters $K_1^{\rm TM}(0)$, $C_1$, and $C_2$,
  we refer to the experimental data of YFe$_{11}$Ti \cite{Nikitin}.
  From the fitting to YFe$_{11}$Ti,
  we determined $C_1=-0.263$, $C_2=-0.237$, and $VK_1^{\rm TM}(0)=47.7$ K/2f.u.
  The resulting temperature dependences of $K_1^{\rm TM}(T)$ are shown in FIG. \ref{YFe11Ti}.
  It should be noted that the $K_1^{\rm TM}(0)$ can be determined by first-principles calculations by introducing a spin-orbit interaction to the valence electrons,
  and the $C_1$ and $C_2$, which are essentially material dependent parameters,
  should be obtained from the respective experimental results.
  However, it is difficult to have all of these parameters,
  so this will be a future issue.


  \subsection{Macroscopic Magnetic Properties}

  Similar to the method applied by Yoshioka et al., \cite{Yoshioka_SmFe12}
  the first-order finite temperature perturbation theory is applied to the
  modified effective lowest-$J$ multiplet Hamiltonian,
  the approximate Gibbs free energy $G({\bm M}_{\rm s},T,{\bm B})$ [Eq. (61) as given in Ref. \cite{Yoshioka_SmFe12}]
  for the whole system can be obtained using the Legendre transformation
  for the Helmholtz free energy $F({\bm M}_{\rm s},T)$  [Eq. (62) in Ref. \cite{Yoshioka_SmFe12}] as follows:
  \begin{align}
    G({\bm M}_{\rm s},T,{\bm B}) =&F({\bm M}_{\rm s},T)-V{\bm M}_{\rm s}(T)\cdot{\bm B},\label{eq:Gan}
  \end{align}
    \begin{align}
    F({\bm M}_{\rm s},T)=&\sum_{j=1}^{n_R}\sum_{p=1}^3\left[  k_{p,j}(T)+\sum
      _{q=1}^{\lfloor p/2\rfloor}k_{p,j}^{q}(T)\cos(4q\Phi)\right]\nonumber\\
    &\times\sin^{2p}\Theta+VK_1^{\rm TM}(T)\sin^2\Theta+F({M}_{\rm s}{\bm n}_c,T),\label{eq:Fan}
  \end{align}
  $\Theta$ and $\Phi$ denote the polar and azimuthal angle of
  ${\bm M}_{\rm s}$, respectively,
  ${\bm n}_c$ is the unit vector alongthe $c$-axis,
  $\lfloor p/2\rfloor$ indicates the
  greatest integer of $p/2$, and
  $k_p^{(q)}(T)$ are the anisotropy constants of the $4f$-shells for the $j$-th $R$ ion,
  part of which are given in Eqs. (56) and (57) in Ref. \cite{Yoshioka_SmFe12} and
  all necessary $k_p^{(q)}$ with a four-fold rotational symmetry are mentioned in the Appendix \ref{Sec:ki}.
  ${\bm M}_{\rm s}(T)$ refers to the spontaneous magnetization for the whole system,
  which can be written as:
  \begin{align}
    {\bm M}_{\rm s}(T) &  =
    \left[\frac{1}{V}\sum_{j=1}^{n_R}m_j(T)+M^{\rm TM}(T)\right]{\bm n}_s,
    \label{eq:Magan}%
  \end{align}
  where ${\bm n}_{\rm s}$ is the direction vector of ${\bm M}_{\rm s}$ and
  $m_j(T)$ is the expectation value of the magnetic moment of the
  $4f$-shell for the $j$-th $R$ ion, which are given
  in Eq. (46) in the Ref. \cite{Yoshioka_SmFe12}.

  The equilibrium condition of the system for a given $T$ and $\bm B$ is:
  \begin{align}
    G^{\mathrm{eq}}(T,{\bm B})&=\min_{{\bm n}_{\rm s}}G({\bm M}_{\rm s},T,{\bm B}),\label{Geq}%
  \end{align}
  In practice, we determine the minimal $G({\bm M}_s,T,{\bm B})$
  numerically by changing the direction of ${\bm M}_{\rm s}$.
  Magnetization curves along the direction of an applied field are obtained using
  $\partial G^{\rm eq}(T, {\bm B})/\partial B$.


  \section{Mechanism of MA in Rare-Earth Ions}
  In this section, we first discuss the mechanism underlying MA induced by $R$ ions in the $R$-transition-metal intermetallic compounds.
  Next, the relationship between the atomic configuration and MA in SmFe$_{12}$ is reviewed.

  \subsection{General Consideration}
  In $R$ permanent magnet materials, the MA of $R$ ions plays an important role.
  An MA is mainly determined by the CF acting on the $4f$ electron cloud in the $R$ ions.
  This CF at the $R$ site is determined by the valence electron cloud surrounding
  the valence electrons and the screened nuclear charges of the ligand ions.
  The contribution of each of the $4f$ electrons and screened nuclear charges to the electrostatic potential is
  sometimes called the valence and lattice contribution, respectively.
  However, it is difficult to clearly separate these contributions in intermetallic compounds.
  Here, we refer to each contribution as the on-site and off-site contribution,
  which is defined as the contribution from the charges inside and outside the atomic radius of the rare earths, respectively.

  \begin{figure}[htb]
    \begin{center}
      \includegraphics[width=8.6cm]{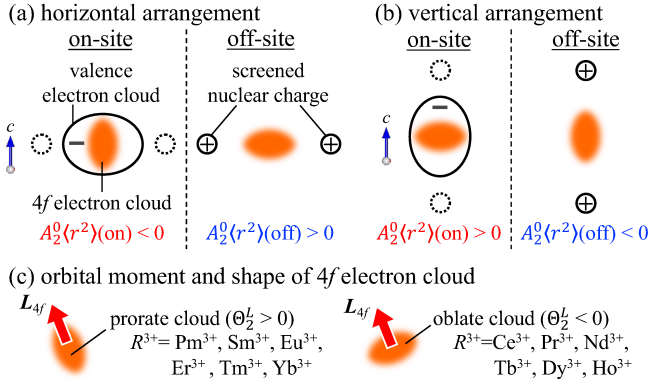} 
      \caption{\label{fig:arrange}
        Schematic of an on-site (left panel) and off-site (right panel) contribution to
        $A_2^0\langle r^2\rangle$
        in (a) horizontal and (b) vertical arrangement of ligands, and
        (c) the orbital moment ${\bm L}_{4f}$ and the shape of the $4f$ electron cloud,
        which are related by the equivalent factor $\Theta^L_2$ in Eq (\ref{eq:eqfct}).
        Ellipses with a minus sign and circles
        with a plus sign represent the valence electron clouds and screened nuclear charges, respectively.}
    \end{center}
  \end{figure}

  We show the effect of its on-site and off-site contributions on MA.
  Within our approximation, the MA constants can be written as a linear combination of the CF parameters,
  which can be decomposed into the on-site and off-site contributions.
  Here, we focus on $A_2^0\langle r^2\rangle$, which is important for MA.
  The left and right panels in FIG. \ref{fig:arrange} (a) and (b) schematically show the stable orientation of the $4f$ electron cloud
  with respect to the CF created by the valence electron cloud and the screened positive charges, respectively.
  The direction of the orbital angular momentum with respect to the $4f$ electron cloud is shown in (c).

  Here, we consider a simple case
  where two ligands are arranged in the (a) horizontal and (b) vertical configuration as shown in the figure.
  When only the on-site contribution is considered,
  the valence electron cloud is oriented in the direction of the ligand,
  and the $4f$ electron cloud is oriented away from it.In contrast, when only the off-site contribution is considered,
  the $4f$ electron cloud is oriented in the direction of the screened nuclear charge.
  Because of the competition between the on-site and off-site contributions,
  the signs of the CF parameters $A_2^0\langle r^2\rangle$(on) and $A_2^0\langle r^2\rangle$(off)
  are opposite, as shown in the figure.
  The sign of $A_2^0\langle r^2\rangle$ is reversed depending on the ligand configuration (a) and (b).

  For typical intermetallic compounds, it has been shown that
  the on-site and off-site contributions compete with each other,
  and the former is dominant, as in the case of SmCo$_5$ \cite{Hummler0} and Nd$_2$Fe$_{14}$B \cite{Hummler1}.
  In these cases, for example, by combining the left panel of (a) and (b) with (c) for SmCo$_5$ and Nd$_2$Fe$_{14}$B, respectively,
  we can understand the mechanism underlying uniaxial MA.
  Consequently, the stable direction of the magnetic moment, on a series of $R^{3+}$,
  is qualitatively determined by the ligand configuration.

  The on-site and off-site contributions to the CF parameters can be quantized by decomposing
  Eq. (\ref{eq:Alm}) into the following form: \cite{Richter2,Hummler0}
  \begin{align}
    A_l^m\langle r^l\rangle({\rm on})=&\frac{a_{l,m}}{2l+1}\frac{e^2}{\varepsilon_0}
    \int_0^{r^R_{\rm AS}}drr^2|R_{4f}(r)|^2\nonumber\\
    &\times\int_{0}^{r^R_{\rm AS}}dr'r'^2\rho_l^m(r')\frac{r_<^l}{r_>^{l+1}},\label{eq:Almon}\\
    A_l^m\langle r^l\rangle({\rm off})=&A_l^m\langle r^l\rangle
    -A_l^m\langle r^l\rangle({\rm on})
    \label{eq:Almoff}
  \end{align}
  with
  \begin{equation}
    \label{eq:rholm}
    \rho_l^m(r)=\int_0^{\pi} d{\theta}\int_0^{2\pi} d{\phi}\sin^2\theta
    \rho_{\rm rest}^0({\bm r})t_l^m(\theta,\phi),
  \end{equation}
  where $\rho({\bm r})$ is the electron density and $r_{<}=\min(r,r')$ and $r_{>}=\max(r,r')$.
  In this study, Eqs (\ref{eq:Almon}) and (\ref{eq:Almoff}) will be referred to as the on-site and off-site contributions, respectively,
  which correspond roughly to the valence and lattice contributions.

  \subsection{MA in the SmFe$_{12}$ compounds}

\begin{figure}[htb]
    \begin{center}
      \includegraphics[width=8.5cm]{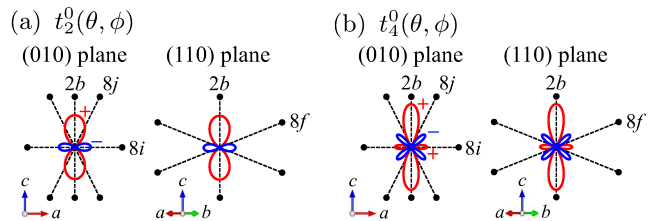} 
      \caption{\label{fig:tlm}
        Tesseral harmonic functions (a) $t_2^0(\theta,\phi)$ and (b) $t_4^0(\theta,\phi)$ in Eq. (\ref{eq:rholm})
        and the Wyckoff positions: $8f$, $8i$, $8j$, and $2b$ in SmFe$_{12}$ on (010) and (110) plane.
      }
    \end{center}
  \end{figure}

  As shown in FIG. \ref{fig:tlm}, in SmFe$_{12}$,
  the nearest Fe is located at the $i$-site and there is a vacancy at the $2b$-site.
  Therefore, the orientation of a $4f$ electron cloud,
  in the shape as shown in the left panel of (a),
  and the development of a uniaxial MA with $K_1>0$ and $K_2<0$
  has been previously discussed,
  qualitatively \cite{Yoshioka_SmFe12}.
  When light elements are doped at the $2b$ site,
  the electronic structure around the Sm ions changes
  and the MA is expected to be greatly affected.
  On the other hand, when Fe is substituted,
  the effect on the MA depends on the replacement site and the substituted atomic species.
  The effects of elemental doping and substitutions
  on MA were investigated as follows.

  \section{Electronic Structure and Model Parameters in SmFe$_{12}$ after Doping and Substitution}
  In this section, we present the results of the first-principle calculations of the electronic structure
  of SmFe$_{12}X$ ($X$=H, B, C, and N) and SmFe$_{11}M$ ($M$=Ti, V, and Co)
  and the model parameters used in the finite temperature calculations.
  Firstly, the partial density of states (PDOS) are shown in FIG. \ref{fig:DosX} and \ref{fig:DosM}.
  The distributions of the charges and magnetic moments are shown in TABLE \ref{tble:chrgX} and \ref{tble:chrgM}.
  Finally, the model parameters used to analyze the magnetic properties
  are shown in TABLE \ref{tble:CFPsX} and \ref{tble:CFPsM},
  where the values of anisotropy constants $k_1$ and $k_2$ on Sm ions at absolute zero are also mentioned.
  To clarify the effect of the screened nuclear charge on the CF,
  the sum of the nuclear and electronic charges in the atomic sphere is shown as the charge at each site.

  \subsection{Model Parameters for SmFe$_{12}X$ ($X$=H, B, C, and N)}
  \begin{figure}[htb]
    \begin{center}
      \includegraphics[width=8.8cm]{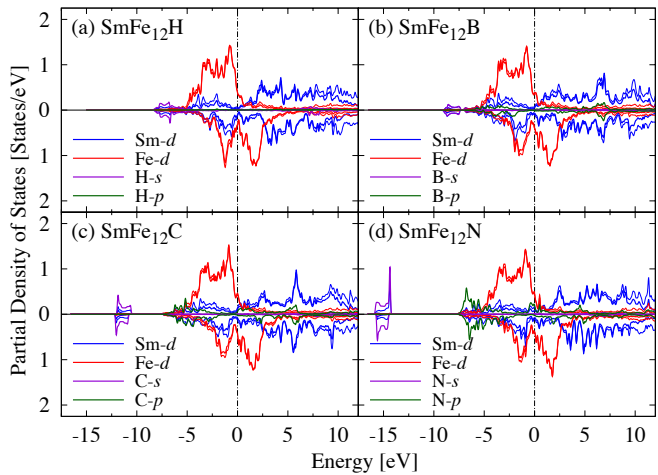}
      \caption{\label{fig:DosX}
        Orbital projected partial density of states (PDOS) in SmFe$_{12}X$ for $X$=(a) H, (b) B, (c) C, and (d) N,
        where the data is averaged for each inequivalent site of Fe ions.
        The PDOS of all orbitals for Sm and Fe are represented by thin solid lines.
        The Fermi level is set to zero.
      }
    \end{center}
  \end{figure}

  The PDOS in SmFe$_{12}X$ ($X$=H, B, C, and N) are shown in FIG. \ref{fig:DosX}.
  The contribution of the $3d$ and $5d$ orbitals are dominant in the PDOS of Fe and Sm ion, respectively.
  Here, the $4f$ electrons in the Sm ion are treated as core electrons, so their contribution does not appear in the PDOS.
  The energy positions of the $s$ and $p$ orbitals of $X$ differ remarkably depending on the type of the light element.
  In the case of H addition, the $1s$ orbital is located at the bottom of the density of states of Fe.
  In the case of B, C, and N doping,
  the weights of the $s$ and $p$ orbitals of the light elements shift to the lower-energy side,
  corresponding to an increase in the nuclear charge.
  For $X$=B and C, strong hybridization of Sm orbitals and $X$-$s$ orbitals is observed around -11 eV and -15 eV, respectively.

  \begin{table*}[htb]
    \caption{\label{tble:chrgX}
      Distribution of charges and magnetic moments on each site
      [Sm($2a$), Fe($8f$), Fe($8i$), Fe($8j$), and $X$(2$b$)]
      and interstitial regions (int.)
      in SmFe$_{12}$ ($M$=n/a) and SmFe$_{12}X$ ($X$=H, B, C, and N) compounds,
      where $e$ and $\mu_{\rm B}$ denote the elementary charge and the Bohr magneton, respectively.
      The partial magnetization excluding the contribution of the $4f$ electrons per two f.u. ($VM^{\rm TM}$) is also shown.
      These results were obtained using the first-principle calculations at absolute zero.
    }
    \begin{ruledtabular}
      \begin{tabular}{ccccccccccccccc}
        &\multicolumn{6}{c}{Charge [$e$]}
        &\multicolumn{7}{c}{Magnetic Moment [$\mu_{\rm B}$]}\\
        \cline{2-7}
        \cline{8-14}
        $X$&
        ${\rm Sm}(2a)$\footnote{Excluding the contribution of $4f$ electrons.}&
        ${\rm Fe}(8f)$&
        ${\rm Fe}(8i)$&
        ${\rm Fe}(8j)$&
        $X(2b)$&
        Int.&
        ${\rm Sm}(2a)$\footnotemark[1]&
        ${\rm Fe}(8f)$&
        ${\rm Fe}(8i)$&
        ${\rm Fe}(8j)$&
        $X(2b)$&
        Int.&
        $VM^{\rm TM}$\footnotemark[1]
        \\ \hline
        n/a  & 5.82 & 1.25 & 1.42 & 1.34 & ----- &-33.78 & -0.50 & 1.90  & 2.56 & 2.32  & ----- & -1.58 & 51.63 \\
        H    & 5.74 & 1.26 & 1.43 & 1.32 & 0.26  &-34.05 & -0.50 & 1.99  & 2.55 & 2.16  & -0.02 & -1.46 & 51.07 \\
        B    & 5.41 & 1.26 & 1.40 & 1.17 & 2.00  &-35.44 & -0.50 & 2.21  & 2.55 & 1.86  & -0.08 & -1.65 & 50.18  \\
        C    & 5.41 & 1.25 & 1.40 & 1.18 & 2.13  &-35.80 & -0.45 & 2.29  & 2.56 & 1.84  & -0.10 & -1.54 & 50.85 \\
        N    & 5.44 & 1.25 & 1.40 & 1.21 & 2.00  &-35.79 & -0.36 & 2.32  & 2.59 & 2.11  &  0.08 & -1.37 & 54.19 \\
      \end{tabular}
    \end{ruledtabular}
  \end{table*}

  The results of doping and substitution on the electronic structure of SmFe$_{12}X$ ($X$=B, C, and N) are shown in TABLE \ref{tble:chrgX}.
  Firstly, in the material SmFe$_{12}$,
  the value of the charge and magnetic moment
  differs depending on the Fe inequivalent site.
  The magnetic moments of Fe($8i$) and Fe($8f$) exhibit the largest and smallest value, respectively.
  These relations are the same as those obtained in a previous study for NdFe$_{12}$ \cite{Miyake}.
  The magnetic moment acting on an Sm ion, excluding the $4f$ electrons, is mainly contributed by $5d$ orbitals,
  which has the opposite sign to that of Fe.
  Doping with a light element at the $2b$ site causes
  the valence electrons bound by the $X$ ions to be distributed in Sm($2a$) and Fe($8j$) adjacent to the $2b$ site and in the interstitial region.
  Thus, it is confirmed that the positive charges of Sm($2a$) and Fe($8j$) decrease.
  Among them, the change in the electron density distribution is small when $X$=H.
  The partial magnetization $VM^{\rm TM}$ increases only in the case of nitrogenation.
  In this case, compared to $X$=B and C,
    the antibonding $p$-orbitals are filled in the up-spin state as shown in FIG. \ref{fig:DosX} (d),
    which leads to a sudden increase in partial magnetization $VM^{\rm TM}$,
    where the magnetic moment of Fe($8j$) adjacent to N($2b$) decreases,
    while the magnetic moment of Fe($8f$) increases significantly.
    This scenario has been discussed in detail using the simplified model by Harashima $et$ $al$ \cite{Harashima2} and
    the results are consistent with those of a previous study for NdFe$_{12}$N compounds \cite{Miyake}.

  TABLE \ref{tble:CFPsX} shows the values of the CF parameters and
  the exchange field $B_{\rm ex}$ acting on the $4f$ shell in the Sm ions of SmFe$_{12}X$ ($X$=H, B, C and N).
  In all cases, the contribution of $A_2^0\langle r^2\rangle$ is dominant.
  Therefore, from the sign of $A_2^0\langle r^2\rangle$,
  it is qualitatively expected that
  the uniaxial and in-plane MA occurs in $X$=H and $X$=B, C, and N, respectively.
  The values of $A_2^0\langle r^2\rangle$=-25 K \cite{KuzminDy}, -65 K \cite{Harashima1}, and -32 K \cite{Delange} were obtained in previous studies.
Among them, -65 K obtained by the same opencore method as ours is the closest value.
On the other hand, in the case of SmFe12N, our result of $A_2^0\langle r^2\rangle$ is 774.7 K, which is larger than the value of 244 K \cite{Harashima1} and 249 K \cite{Delange}.
This discrepancy is thought to be caused by the lack of structural optimization in our calculations.
Harashima et al. show that the volume expands due to nitriding.
By taking this strain into account in our calculations, the hybridization of Sm and N sites is weakened, 
and the value of the CF $A_2^0\langle r^2\rangle$ is expected to be smaller.
  Each value of $B_{\rm ex}(0)$ is not related to the Fe spin density,
  but is roughly proportional to the valence electron spin density at the Sm site shown in TABLE \ref{tble:chrgX}.
  This is consistent with the mechanism
  proposed by Brooks $et$ $al$. \cite{Brooks} and examined by Liebs $et$ $al$. \cite{Liebs2},
  where the exchange field originates from the intra-atomic $4f$-$5d$ coupling
  and no major contribution of non-local $4f$-$3d$ effective interaction persists, hence,
  our estimation of $B_{\rm ex}(0)$ is justified.
  From these parameters, the MA constants
  $k_1(0)$ and $k_2^{(1)}(0)$ per  Sm ion can be obtained analytically.
  The results are shown in columns 8 and 9.
  The MA constant $k_1(0)$ differs depending on the doped element.
  Particularly, in the case of hydrogenation,
  the value of $k_1(0)$ becomes about twice larger than that of SmFe$_{12}$.

  \begin{table}[htb]
    \caption{\label{tble:CFPsX}
      CF parameters $A_l^m\langle r^l\rangle$ [K] for ($l, m$) and
      exchange field $B'_{\rm ex}=\mu_{\rm B}B_{\rm ex}(0)/k_{\rm B}$ [K]
      obtained using the first-principle calculations
      in SmFe$_{12}$ ($M$=n/a) and SmFe$_{12}X$ ($X$=H, B, C, and N) compounds,
      where $\mu_{\rm B}$ and $k_{\rm B}$ are the Bohr magneton and Boltzmann constant, respectively.
      First(Second) order MA constant for a single Sm ion $k_{1,2}^{(1)}(0)$ [K] at $T=0$ K is also shown,
      which were analytically obtained using Eqs. (56) and (57) as given in Ref. \cite{Yoshioka_SmFe12}.
    }
    \begin{ruledtabular}
      \begin{tabular}{ccccccccccccccccccc}
        $X$&
        (2,0)&
        (4,0)&
        (4,4)&
        (6,0)&
        (6,4)&
        $B'_{\rm ex}$&
        $k_1$&
        $k_2$&
        $k_2^1$
        \\ \hline
        n/a\footnote{Previous calculaion shown in Ref. \cite{Yoshioka_SmFe12}.}
        & -71.4  & -21.3 & -49.3 & 5.9 & 3.0 &296.1&   97.7 & -40.9 & -2.3\\
        H   & -163.7 & -49.5 & -6.7 & 3.9 & 3.2  &294.3&  220.1 & -83.5 & -0.4\\
        B   &  172.2 & -31.4 & 17.9 &-13.4& 8.6  &292.3&  -84.4 & -34.1 & 0.5 \\
        C   &  439.0 & -2.3  & 27.9 & -8.1& 12.0 &261.3& -343.8 & 6.0   & 0.9 \\
        N   &  774.7 & 48.1  & 29.8 & 2.1 & 12.2 &203.2& -683.5 & 72.2  & 1.0
      \end{tabular}
    \end{ruledtabular}
  \end{table}

  \subsection{Mechanism of MA in SmFe$_{12}X$ ($X$=H, B, C, and N)}\label{sec:originX}
\begin{figure}[htb]
    \begin{center}
      \includegraphics[width=8.8cm]{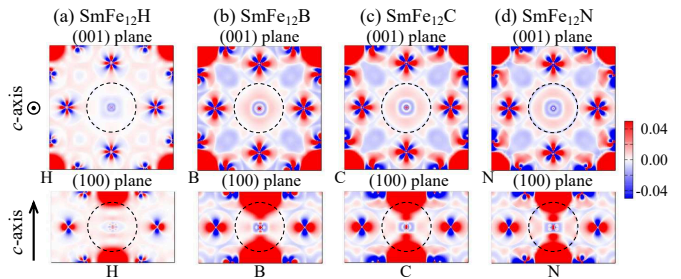}
      \caption{\label{fig:rhodiffX}
        Difference in the electron density $\Delta \rho^0_{\rm rest}({\bm r})$ [$a_0^{-3}$]
        between SmFe$_{12}X$ [$X$=(a) H, (b) B, and (c) N] and SmFe$_{12}$,
        where $a_0$ is the Bohr radius.
        An Sm ion is located at the center of the figure and the
        dashed circles represent the atomic sphere radius $r^{\rm Sm}_{\rm AS}=3.2 a_0$.
      }
    \end{center}
  \end{figure}

  FIG. \ref{fig:rhodiffX} shows the change in the charge density owing to the doping of light elements at the
  $2b$-site,
  where an Sm ion is located in the middle of the figure.
  $\Delta\rho({\bm r})$ is defined as the charge density of SmFe$_{12}X$ minus that of SmFe$_{12}$.
  The common feature of the two is the increase in the electron density around the $2b$-site
  due to the addition of a light element $X$.
  This increased electron density gets distributed
  inside the atomic sphere of Sm because the binding of the added element $X$ is weak.
  As shown in the lower panel, in the case of (a) $X$=H, there is no significant change in the charge density near the Sm site.
  On the other hand, in the case of (b)-(d) $X$=B, C, and N,
  the change extends to the vicinity of the Sm site.

\begin{figure}[htb]
    \begin{center}
      \includegraphics[width=8.0cm]{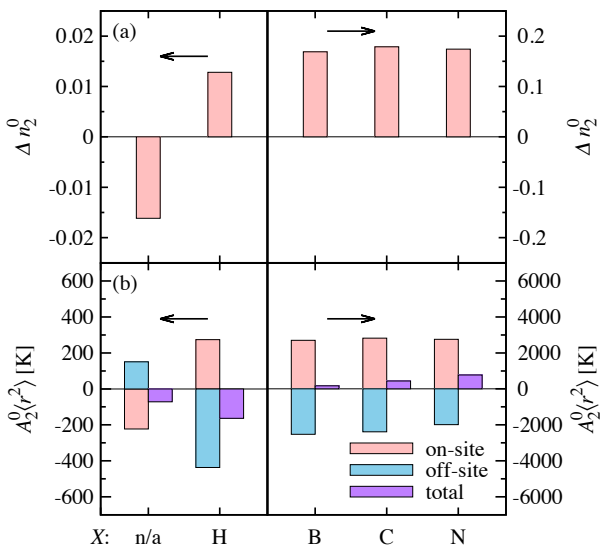} 
      \caption{\label{fig:AlmX}
        (a) Asphericity of valence electron cloud defined by
        $\Delta n_2^0=\int_0^{r^{\rm Sm}_{\rm AS}}r^2\rho_2^0(r)dr$
        and (b) on-site and off-site contribution to $A_2^0\langle r^2\rangle$
        on Sm site in SmFe$_{12}$ ($X$=n/a) and SmFe$_{12}X$ ($X$=H, B, C, and N).}
    \end{center}
  \end{figure}

  To clarify the effect of this charge density, change on the CF,
  FIG. \ref{fig:AlmX} (a) shows the distribution $\Delta n_2^0$ of the electron density in the Sm $2a$-site,
  which is defined by $\Delta n_2^0=\int_0^{r^{\rm Sm}_{\rm AS}}r^2\rho_2^0(r)dr$.
  Compared to the case of SmFe$_{12}$, the $\Delta n_2^0$ changes to the opposite sign with the addition of light elements,
  indicating that the valence electrons on the Sm site are biased in the prolate shape along the $c$-axis.
  Especially in the case of $X$=B, C, and N, the value of $\Delta n_2^0$ is one order of magnitude larger.

  The CF parameters decomposed into the on-site and off-site contributions (FIG. \ref{fig:AlmX} (b))
  show that there is a strong correlation between the asphericity of charge density and $A_2^0\langle r^2\rangle$(on).
  In the case of SmFe$_{12}$ $(X=n/a)$,
  the on-site contribution is more dominant than the off-site one,
  indicating the occurrence of uniaxial MA \cite{Yoshioka_SmFe12}.
  Conversely, when a light element is added at the $2b$-site,
  both the on-site and off-site contributions have opposite signs to those of SmFe$_{12}$.
  This reflects that the nearest neighbor ions to the Sm site are switched from the horizontal arrangement [FIG. \ref{fig:arrange} (a)]
    to the vertical arrangement [FIG. \ref{fig:arrange} (b)] on the addition of the light elements.
  Comparing the magnitude of the on-site and off-site contributions for each substitutional element,
  the off-site contribution is larger only for $X$=H.
  When H is added to the $2b$-site, the change in the charge density near Sm is small
  and is largely affected by the positive nuclear charge.
  Therefore, the off-site contribution becomes dominant and $A_2^0\langle r^2\rangle$(on) becomes negative.
  Conversely, in the case of $X$=B, C, and N, the change in the charge density
  just above the Sm site is large and the valence contribution
  becomes more dominant than the lattice contribution, and the sign of $A_2^0\langle r^2\rangle$ becomes positive.
  In summary, for $X$=H the lattice contribution is dominant and exhibits a strong uniaxial MA,
  while for $X$=B, C, and N, the valence contribution is dominant and displays a strong in-plane anisotropy.

  \subsection{Model Parameters for SmFe$_{11}M$ ($M$=Ti, V, and Co)}\label{sec:originM}

  \begin{figure*}[htb]
    \begin{center}
      \includegraphics[width=13.0cm]{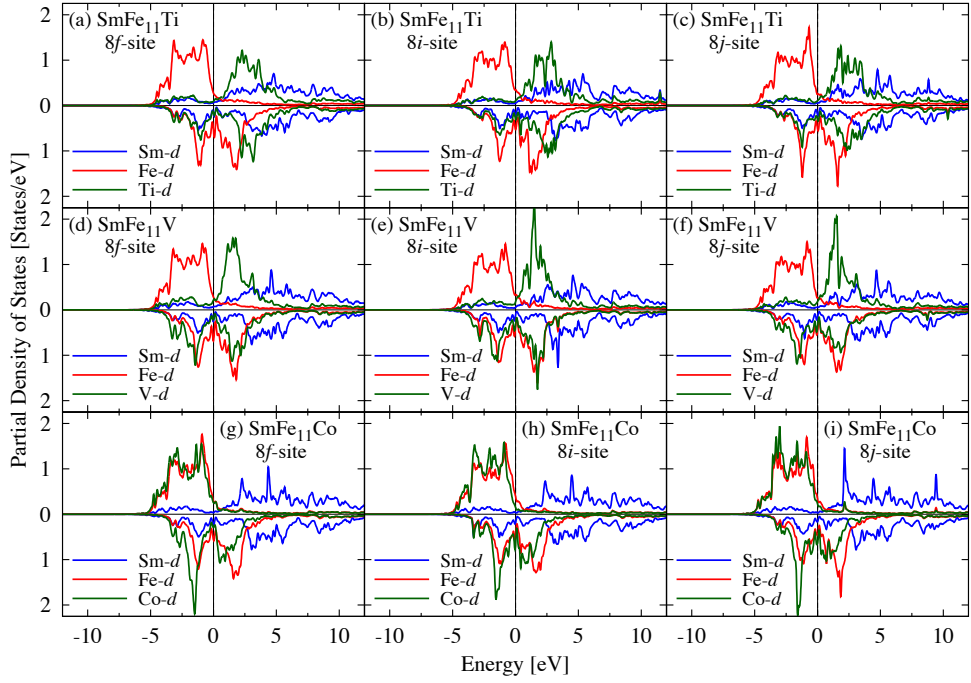}
      \caption{\label{fig:DosM}
        Orbital projected partial density of states in SmFe$_{11}M$ for $M$=(a-c) Ti, (d-f) V, and (g-i) Co for each replacement site,
        where the data is averaged for each ion. The Fermi level is set to zero.
      }
    \end{center}
  \end{figure*}
  The PDOS for SmFe$_{11}M$ ($M$=Ti, V, and Co) is shown in FIG. \ref{fig:DosM}.
  While the averaged PDOS for the inequivalent sites of Fe show a similar distribution,
  the PDOS for the substitutional elements are characteristic in their nature.
  In the case of $M$=Ti and V, the magnetic polarization is in a direction opposite to that of Fe,
  and in the case of $M$=Co, the polarization is in the same direction as that of Fe.
  In the case of Co substitution, the number of occupied electrons, especially in the minority band, increases.
  From these results, Ti and V are expected to have opposite magnetic moments with respect to Fe,
  while Co is parallel to Fe but with a decreasing magnitude.

  \begin{table*}[htb]
    \caption{\label{tble:chrgM}
      Same as TABLE \ref{tble:chrgX} but in SmFe$_{11}M$ ($M$=Ti, V, and Co)
      for each replacement site,
      which is indicated by the subscript to $M$.
      The columns of TM1, TM2, and TM3 show the averaged magnetic moments
      at the $8f$, $8i$, and $8j$ sites in the original SmFe$_{12}$ and
      numbers in parenthesis denote the magnetic moment of the $M$ ion.
    }
    \begin{ruledtabular}
      \begin{tabular}{cccccccccccc}
        &\multicolumn{5}{c}{Charge [$e$]}
        &\multicolumn{6}{c}{Magnetic Moment [$\mu_{\rm B}$]}\\
        \cline{2-6}
        \cline{7-12}
        $M$&
        ${\rm Sm}$\footnote{Excluding the contribution of the $4f$ electrons.}&
        TM1&
        TM2&
        TM3&
        Int.&
        ${\rm Sm}$\footnotemark[1]&
        TM1&
        TM2&
        TM3&
        Int.&
        $VM^{\rm TM}$\footnotemark[1]
        \\ \hline
        n/a         & 5.82 & 1.25        & 1.42        & 1.34        & -33.78 & -0.50 & 1.90        & 2.56        & 2.32        & -1.58 & 51.63 \\
        Ti$_{8f}$   & 5.91 & 1.41(1.80)  & 1.45        & 1.37        & -35.69 & -0.52 & 1.30(-0.70) & 2.49        & 2.34        & -2.09 & 45.85 \\
        Ti$_{8i}$   & 5.94 & 1.28        & 1.58(1.97)  & 1.38        & -35.79 & -0.55 & 1.87        & 1.69(-0.72) & 2.28        & -2.21 & 43.39 \\
        Ti$_{8j}$   & 5.90 & 1.27        & 1.45        & 1.52(1.90)  & -35.74 & -0.55 & 1.90        & 2.48        & 1.59(-0.69) & -2.18 & 44.46 \\
        V$_{8f}$    & 5.91 & 1.39(1.72)  & 1.45        & 1.37        & -35.49 & -0.50 & 1.17(-1.27) & 2.47        & 2.32        & -2.06 & 44.64 \\
        V$_{8i}$    & 5.92 & 1.28        & 1.55(1.87) & 1.38        & -35.52 & -0.54 & 1.87        & 1.48(-1.45) & 2.26        & -2.18 & 41.64 \\
        V$_{8j}$    & 5.90 & 1.27        & 1.45        & 1.49(1.80)  & -35.49 & -0.54 & 1.91        & 2.45        & 1.39(-1.39) & -2.18 & 42.75 \\
        Co$_{8f}$   & 5.87 & 1.24(1.15)  & 1.44        & 1.37        & -34.12 & -0.51 & 1.93(1.53)  & 2.60        & 2.44        & -1.67 & 53.13 \\
        Co$_{8i}$   & 5.87 & 1.27        & 1.41(1.30)  & 1.37        & -34.09 & -0.52 & 1.98        & 2.37(1.74)  & 2.37        & -1.66 & 51.06 \\
        Co$_{8j}$   & 5.87 & 1.27        & 1.44        & 1.33(1.23)  & -34.11 & -0.52 & 2.05        & 2.61        & 2.19(1.62)  & -1.69 & 52.10 \\
      \end{tabular}
    \end{ruledtabular}
  \end{table*}

  Next, the charge and magnetic properties of SmFe$_{11}M$ ($M$=Ti, V, and Co) is reported in TABLE \ref{tble:chrgM}.
  Here, the numbers in parentheses denote the magnitude of the charge and spin moments of the substitution sites.
  First, we note the charge distribution.
  Regardless of the location of the substitution, when $M$=Ti or V, electron binding at the substitution site is weakened.
  As a result, the amount of charge at the substitution site increases,
  and the electron density in the interstitial region increases.
  On the contrary, when $M$=Co, the amount of charge on the substitution site decreases slightly because Co attracts more electrons.
  Next, we focus on the magnetic moment.
  In the case of $M$=Ti or V, the substitution element has a magnetic moment that is antiparallel to Fe.
  As a result, the partial magnetization $VM^{\rm TM}$ is greatly reduced.
  On the other hand, when Fe is replaced by Co, the magnetic moment of Co itself decreases as indicated in the parenthesis,
  but the magnitude of the partial magnetization $VM^{\rm TM}$ increases or decreases depending on the substitution site,
  but the change is smaller than as that of element doping.

  Similar to the case of element doping,
  the values of $A_l^m\langle r^l\rangle$, $B_{\rm ex}(0)$ and the $k_{1,2}^{(1)}(0)$
  in SmFe$_{11}M$ ($M$=Ti, V, and Co) are shown in TABLE \ref{tble:CFPsM}.
  The CF parameters in each $M$ exhibit variations depending on the substitution site even for the same substitution element.
  In particular, the value of $A_2^0\langle r^2\rangle$ is positive only
  when the Fe($8i$) site is replaced by Ti.
This has been confirmed in the previous study \cite{Harashima1}, 
which showed that the Fe($8i$) site is preferentially replaced by Ti and the value of $A_2^0\langle r^2\rangle$=8 K. 
This is consistent with our results in that it takes a small positive value.
  On the other hand, the value of $B_{\rm ex}$ is almost the same regardless of the type of substitution, owing to the comparable spin moments of Sm (shown in TABLE II)
  .
  Based on the values of the MA constants obtained from these parameters,
  $k_1$ is always positive, regardless of the type of substitution.
  The MA is significantly increased by substituting Ti and V with Fe($8j$),
  and in the case of Co substitution,
  the MA increases  irrespective of the location of the substituted site.
  However, when Fe($8i$) is replaced by Ti,
  the uniaxial MA is greatly reduced because $k_2(0)$ has a larger negative magnitude than $k_1(0)$.
  Note that for $M$=Ti$_{8i}$, $A_4^4\langle r^4\rangle$ possesses a large value of -112.7 K.
  This contributes to the in-plane MA through $k_2^1$, as shown in Eq. \ref{eq:k21},
  however, due to the small $A_4^4\langle r^4\rangle$ coefficient, the value of $k_2^1$ is small.

  \begin{table}[htb]
    \caption{\label{tble:CFPsM}
      Same as TABLE \ref{tble:CFPsX} but in SmFe$_{11}M$ ($M$=Ti, V, and Co) for each replacement site,
      which is indicated by the subscript to $M$.
      CF parameters caused by local rotational symmetry breaking
      due to elemental substitution are shown in TABLE \ref{tble:CFPres} in Appendix \ref{Sec:Aprm}.
    }
    \begin{ruledtabular}
      \begin{tabular}{ccccccccccc}
        $M$&
        (2,0)&
        (4,0)&
        (4,4)&
        (6,0)&
        (6,4)&
        $B'_{\rm ex}$&
        $k_1$&
        $k_2$&
        $k_2^1$
        \\ \hline
        n/a\footnote{Previous calculations are presented in the Ref. \cite{Yoshioka_SmFe12}.}
        & -71.4 & -21.3 & -49.3  & 5.9  & 3.0   & 296.1 & 97.7 & -40.9 & -2.3 \\
        Ti$_{8f}$  & -49.8 & -21.3 & 48.7   & 5.0  & -3.2  & 301.5 & 80.2 & -39.8 & 2.3  \\
        Ti$_{8i}$  & 19.9  & -34.5 & -112.7 & 6.0  & -0.8  & 319.5 & 49.4 & -61.0 & -5.0 \\
        Ti$_{8j}$  &-212.4 & -13.6 & -33.5  & 6.4  & -17.3 & 319.6 & 192.5& -28.7 & -1.0 \\
        V$_{8f}$   & -57.1 & -25.6 & 32.5   & 5.2  & -5.8  & 285.1 & 94.2 & -47.2 & 1.7  \\
        V$_{8i}$   & -71.7 & -29.5 & -26.5  & 5.8  & -4.6  & 317.1 & 111.7& -53.1 & -1.0 \\
        V$_{8j}$   &-211.2 & -18.2 & 18.2   & 6.1  & -17.0 & 316.6 & 199.9& -35.6 & 1.3  \\
        Co$_{8f}$  &-78.9  & -22.6 & -32.7  & 5.9  & -0.5  & 301.4 & 105.7& -42.9 & -1.5 \\
        Co$_{8i}$  &-120.4 & -18.0 & 4.2    & 5.5  & -2.4  & 304.5 & 129.4& -35.0 & 0.3  \\
        Co$_{8j}$  &-94.3  & -23.1 & -25.8  & 5.4  & 2.1   & 306.3 & 118.1& -42.9 & -1.2 \\
      \end{tabular}
    \end{ruledtabular}
  \end{table}

  \subsection{Mechanism of MA in SmFe$_{11}M$ ($M$=Ti, V, and Co)}
\begin{figure}[htb]
    \begin{center}
      \includegraphics[width=8.8cm]{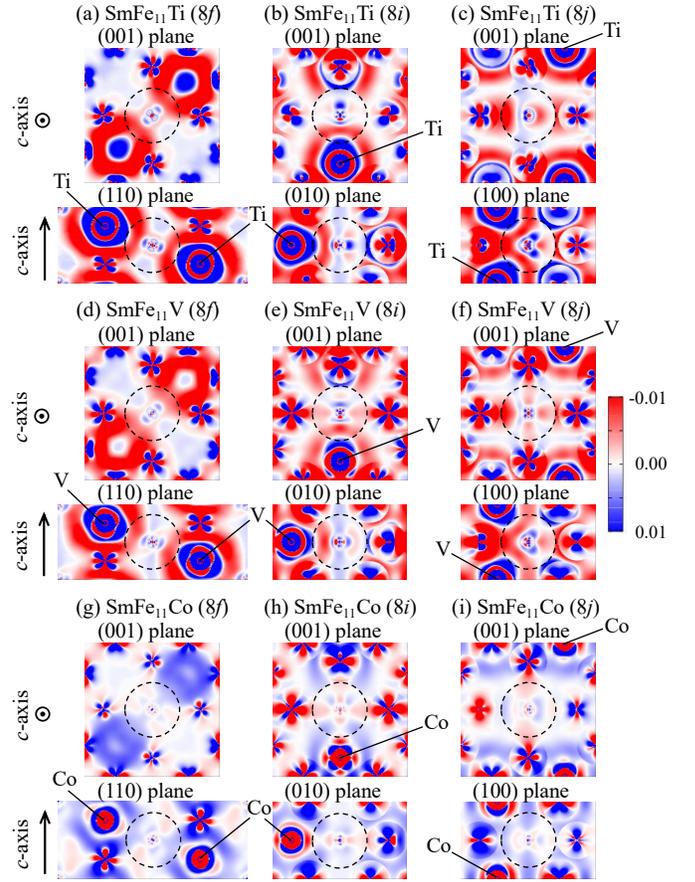}
      \caption{\label{fig:rhodiffM}
        Difference in the electron density $\Delta \rho^0_{\rm rest}({\bm r})$ [$a_0^{-3}$]
        and screened nuclear charge [$e$], which is shown below the symbols $M$,
        between SmFe$_{11}M$ [$M$=(a)-(c) Ti, (d)-(f) V, and (g-i) Co] and
        SmFe$_{12}$ with substitution for each Fe $8f$, $8i$, and $8j$ site in SmFe$_{12}$.
        An Sm ion is located at the center of the figures and the
        dashed circles represent the atomic sphere radius $r^{\rm Sm}_{\rm AS}=3.2 a_0$.
        In order to measure the difference, the crystal structure of SmFe$_{12}$ is adjusted to each one of SmFe$_{11}M$.
      }
    \end{center}
  \end{figure}

  FIG. \ref{fig:rhodiffM} shows the change in the charge density after elemental substitution.
  The substitution increases or decreases the screened nuclear charge and changes the charge density in the other regions.
  The screened nuclear charge acts directly on the $4f$ electron cloud as an electrostatic potential.
  Thus, a change in the screened nuclear charge at the substitution site is also important.
  The charges at the $8f$, $8i$, and $8j$ sites of Fe in SmFe$_{12}$ are 1.25, 1.42, and 1.34, respectively.
  The change in the screened nuclear charge of the substituted ions is shown by the numbers under the $X$ symbol.

\begin{figure}[htb]
    \begin{center}
      \includegraphics[width=8.0cm]{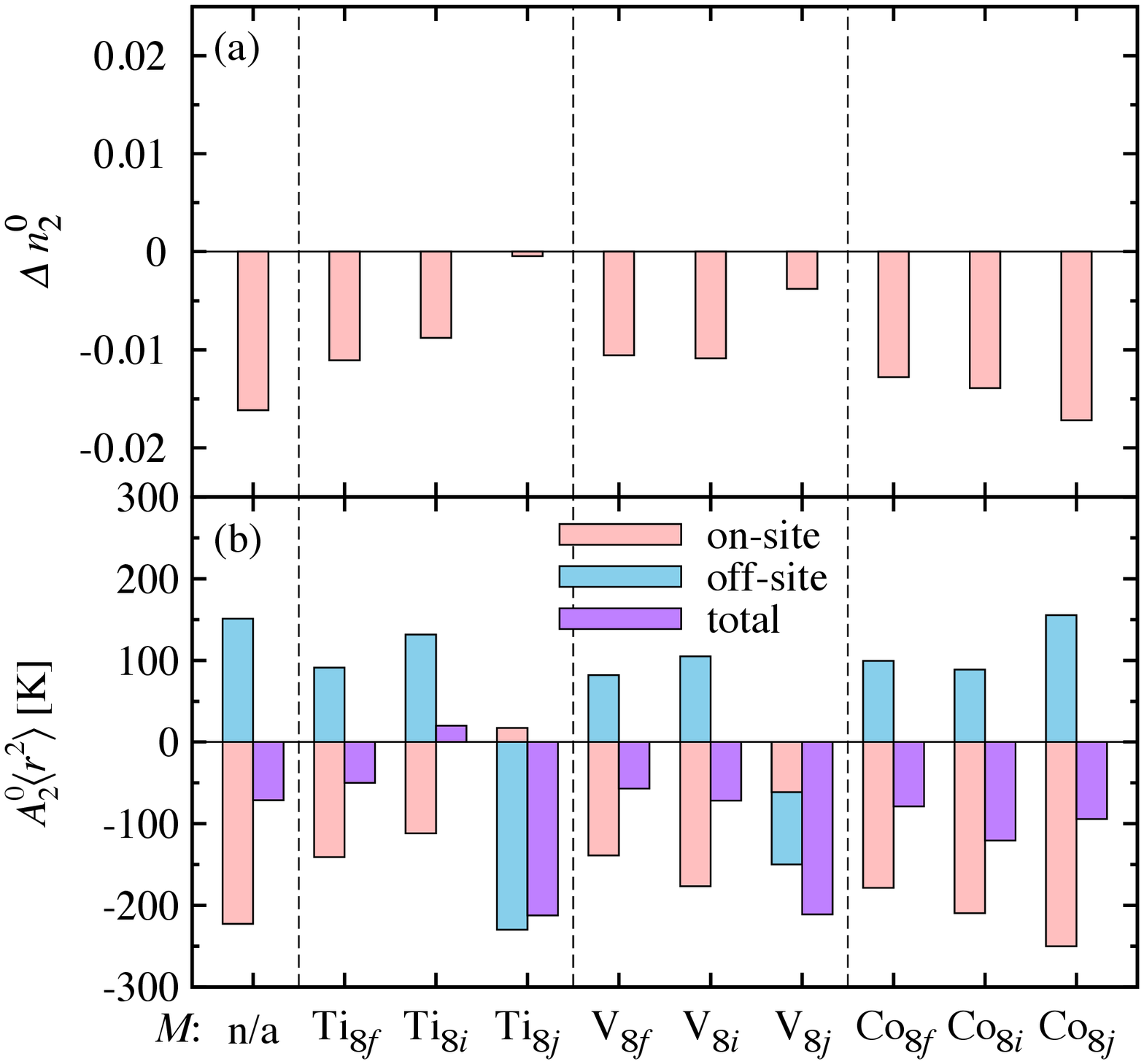} 
      \caption{\label{fig:AlmM}
        (a) Asphericity of valence electron cloud defined by
        $\Delta n_2^0=\int_0^{r^{\rm Sm}_{\rm AS}}r^2\rho_2^0(r)dr$
        and (b) on-site and off-site contribution to $A_2^0\langle r^2\rangle$
        at the Sm site in SmFe$_{12}$ ($M$=n/a) and SmFe$_{11}M$ ($M$=Ti, V, and Co)
        for each replacement site,
        which is indicated by the subscript to $M$.}
    \end{center}
  \end{figure}

  Since the charge density difference $\Delta \rho({\bm r})$ shows a complicated distribution,
  it is shown together with the distribution parameter shown in FIG. \ref{fig:AlmM} (a).
  Originally, the valence electrons at the Sm site have an oblate shape.
  Since the sign of $\Delta n_2^0$ does not change with elemental substitution,
  it can be seen from FIG. \ref{fig:AlmM} (a) that an oblate shape similar to the one in  SmFe$_{12}$ is realized.

  It can be seen in FIG. \ref{fig:rhodiffM} that both
  the positive nuclear charge at each substitution site and the electron density
  increases for a Ti and V substitution and decreases for a Co substitution.
  However, since the charge density difference shows a complicated distribution,
  we confirm the asphericity of the valence electron cloud in $\Delta n_2^0$ as shown in FIG. \ref{fig:AlmM} (a).
  Originally, the valence electron of the Sm site in SmFe$_{12}$ has an oblate shape ($\Delta n_2^0<0$).
  Since the sign of $\Delta n_2^0$ does not change even after an elemental substitution,
  a similar oblate shape of the valence electron cloud as that of SmFe$_{12}$ is realized.

  In the following section, we discuss the results of a detailed investigation of
  the substitution of light i.e., Ti and V and heavy i.e., Co elements, separately.
  As shown in FIG. \ref{fig:rhodiffM}, when Fe is substituted for Ti and V,
  the weak positive charge of the nucleus weakens the electron binding,
  resulting in an increase in the screened nuclear charge and the electron density in the interstitial region.
  Consequently, the asphericity decreases in both the cases of Ti and V substitution.
  In particular, when Fe($8j$) is substituted for Ti and V, $\Delta n_2^0$ decreases significantly.
  This is probably due to the increase in the charge density near the $8j$ substitution site,
  as shown in the lower panels of FIG. \ref{fig:rhodiffM} (c) and (f).
  Next, we focus on the effect of the charges outside the atomic sphere on $A_2^0\langle r^2\rangle$(off-site)
  and clarify the mechanism that determines the total $A_2^0\langle r^2\rangle$.
  When the Fe at the $8j$ site is replaced by Ti or V, the positive charge at the site increases,
  and the $4f$ electrons of Sm shown in the right panel of FIG. \ref{fig:arrange} (b) are
  stabilized in a prolate shape with respect to the $c$-axis.
  In fact, from the configuration between the $8j$ site of Sm, as shown in FIG. \ref{fig:tlm} (a),
  the off-site contribution $A_2^0\langle r^2\rangle$(off)
  and the total $A_2^0\langle r^2\rangle$ attains a large negative magnitude,
  as shown in FIG. \ref{fig:AlmM} (b).
  However, when one of the $8i$-site adjacent to Sm is substituted with Ti or V,
  the off-site contribution is positively larger than that of the other substitution sites.
  Especially in the case of Ti substitution,
  where the increase in the positive charge is large, the prolate orientation
  of $4f$ electrons is stabilized.
  This is reflected in the positive value of the total $A_2^0\langle r^2\rangle$, as shown in FIG. \ref{fig:AlmM} (b).

  In the case of Co substitution, the positive charge of the screened nucleus and the electron density in the interstitial region
  decrease because of the attraction of the large positive charge of Co ions to electrons.
  In particular, the electron density in the $c$-axis direction
  from Sm to the $2b$ site, in the presence of a void, is reduced regardless of the substitution site.
  This can be confirmed by the profiles of the lower panel depicted in FIG. \ref{fig:rhodiffM} (g), (h), and (i).
  Hence, value of the total $A_2^0\langle r^2\rangle$ decreases as shown in FIG. \ref{fig:AlmM} (b).
  As a result, the prolate $4f$ electron cloud tends to strongly fix along the $c$-axis direction,
  and subsequently the MA increases.

  When Fe($8j$) located above and along the $c$-axis of Sm is replaced by Ti or V, the MA  greatly increases owing to the effect of the shielded positive charge.
  Conversely, when Fe($8i$) located laterally along the $a$ or $b$-axis of Sm is replaced by Ti, the MA changes from uniaxial to in-plane due to the influence of the screened positive charge.
  When Fe is replaced by Co, the MA increases independent of the substitution site.
  This occurs owing to a decrease in the charge density along the $c$-axis from Sm and the fixing of the Sm $4f$ electron cloud along the direction of the $c$-axis.

  \section{Changes in the Macroscopic Magnetic Properties in SmFe$_{12}$ as a Result of Doping and Substitution}

  The changes in the magnetic properties of SmFe$_{12}X$ ($X$=H, B, C, and N) and SmFe$_{11}M$ ($M$=Ti, V, and Co)
  are discussed in subsections A and B, respectively.
  The MA of SmFe$_{12}$ is observed to change drastically as a result of doping and substitution, as summarized in TABLE \ref{tble:CFPsX} and \ref{tble:CFPsM}.
  In order to clarify the cause of the variation,
  the change in the charge distribution due to doping and substitution is illustrated,
  and the results of decomposing the CF $A_2^0\langle r^2\rangle$ of Sm into on-site and off-site contributions are also presented.
  On the other hand, the temperature dependence of the magnetization is mainly determined by the transition-metal elements,
    which accounts for most of the magnetization.
  In the present study, this is treated phenomenologically except for the values at absolute zero,
  and the temperature dependence of the magnetization of the whole system is not considerably different from FIG. \ref{YFe11Ti}.
  In fact, the magnetic moment carried by the 4f electrons in Sm is as small as 0.3 $\mu_B$ per atom \cite{Yoshioka_SmFe12}.
  Finally, the MA constants and magnetization curves at finite temperatures are shown.

  \subsection{MA Constants and Magnetization Curves in SmFe$_{12}X$ ($X$=H, B, C, and N)}

  We clarify the mechanism underlying the MA using the charge density distribution.
  Finally, we show the results of $K_1$ and $K_2$ at finite temperatures.

  \begin{figure}[htb]
    \begin{center}
      \includegraphics[width=7.0cm]{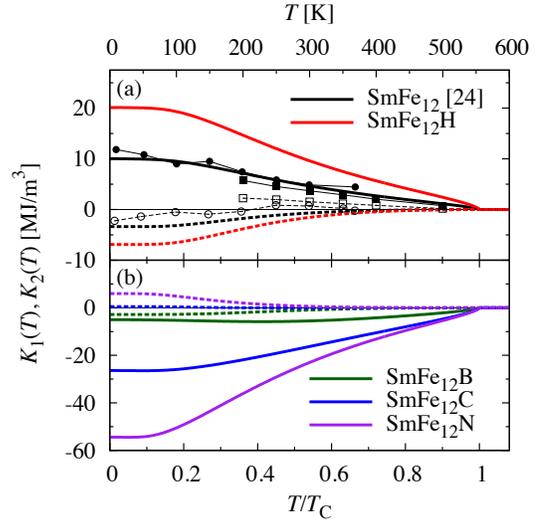}
      \caption{\label{fig:K1K2X}
        Temperature-dependent MA constants $K_1(T)$ (solid lines) and
        $K_2(T)$ (broken lines) [MJ/m$^3$] in (a) SmFe$_{12}$ \cite{Yoshioka_SmFe12} and SmFe$_{12}X$ [$X$=(a) H and (b) B, C, and N] compounds,
        where temperature is scaled using the Curie temperature $T_{\rm C}$.
        These results were obtained by using the analytical method in Eqs. (21) and (22) as mentioned in the Ref. \cite{Yoshioka_SmFe12}.
        The experimental results for $K_1(T)$ and $K_2(T)$ in SmFe$_{12}$ are represented
        by solid and open plots, respectively, where
        the circles and squares represent the results measured in different experiments of the Sucksmith-Thompson method \cite{Hirayama2}
        and the anomalous Hall effect \cite{Ogawa}, respectively.
        Actual value of temperature for SmFe$_{12}$ ($T_{\rm C}=555$ K) is shown on the top of the graph.
      }
    \end{center}
  \end{figure}

  The results for the finite temperature MA constants $K_1$ and $K_2$
  calculated by using the analytical expressions
  in Eqs. (\ref{eq:K1an}) and (\ref{eq:K2an}) are shown in FIG. \ref{fig:K1K2X}.
  Here, the horizontal axis represents the temperature scaled using the Curie temperature $T_{\rm C}$.
  At all temperatures, the absolute value of $K_1$ is larger than that of $K_2$.
  Next, we focus on the temperature dependence.
  According to Eq. (\ref{eq:K1an}), $A_2^0\langle r^2\rangle$ and $A_4^0\langle r^4\rangle$
  contribute to $K_1$ with the same sign,
  and it is known that the terms containing higher order CF parameters with a high number of $l$ decay
  quickly with increasing temperature.
  In the case of SmFe$_{12}$ and SmFe$_{12}X$ of $X$=H and N, $A_2^0\langle r^2\rangle$ and $A_4^0\langle r^4\rangle$
  have the same sign and exhibit a monotonic temperature dependence.
  On the contrary, for SmFe$_{12}X$ of $X$=B and C, they are opposite in signs,
  and especially for SmFe$_{12}$B, they exhibit a non-monotonic temperature dependence.
  This is due to the fact that the $A_4^0\langle r^2\rangle$ term, which contributes positively to $K_1$,
  decays quickly with temperature.
  For $X$=H, the MA is increased by a factor of about 2 in $K_1$,
  which is expected to improve the magnetic properties.
  However, when light elements C, and N are added,
  strong in-plane anisotropy is observed due to the large positive value of $A_2^0\langle r^2\rangle$.

\begin{figure}[htb]
    \begin{center}
      \includegraphics[width=8.5cm]{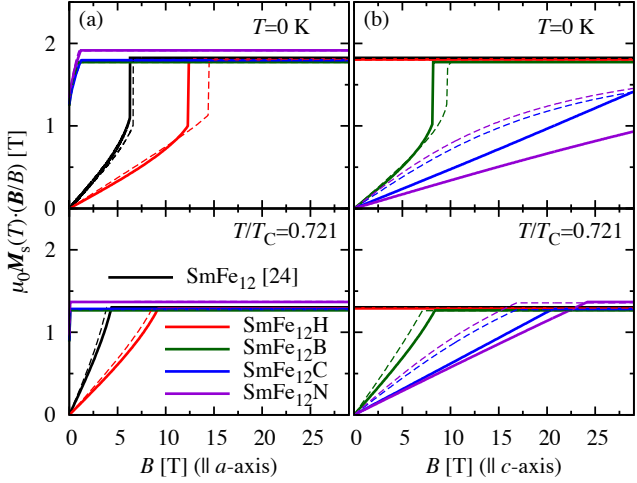}
      \caption{\label{fig:magcurveX}
        Magnetization curves along the (a) $a$-axis and (b) $c$-axis at $T$=0 and $T/T_{\rm C}=0.721$
        in SmFe$_{12}$ and SmFe$_{12}$X [$X$=H, B, C and N]
        compounds.
        These results are obtained by using the analytical method in Eq. (63) mentioned in Ref. \cite{Yoshioka_SmFe12}.
        The thin broken curves show the statistical results obtained from Eq. (25)} given in Ref. \cite{Yoshioka_SmFe12}.
    \end{center}
  \end{figure}

  FIG. \ref{fig:magcurveX} shows the magnetization curve obtained analytically
  within the framework of the linear theory for the CF.
  Here we also show the results calculated by the statistical method
  Using the exact diagonalization, as shown by broken curves.
  (a) and (b) show the results when the magnetic field is applied in the $a$ and $c$-axis directions, respectively.
  (a) and (b) together exhibit a uniaxial MA in the case of SmFe$_{12}$ and SmFe$_{12}$H,
  and an in-plane anisotropy in the case of SmFe$_{12}X$ for $X$=B, C, and N.
  In particular, comparing SmFe$_{12}$ and SmFe$_{12}$H,
  we can see that the MA is enhanced by hydrogenation.
  Moreover, the first-order magnetization process (FOMP) occurs at low temperatures.
  This can be deduced from the competition between the MA constants $K_1(T)$ and $K_2(T)$.
  In fact, the FOMP condition $-K_2(T)<K_1(T)<-6K_2(T)$\cite{Yoshioka_SmFe12} is satisfied in this case.
  The strong in-plane anisotropy is observed for $X$=C and N.
  In this case, the difference between the results calculated by the analytical (solid curves) and
  statistical (broken curves) methods becomes large.
  This means that the framework of the linear approximation for CF is broken
  because the value of the $A_2^0\langle r^2\rangle$ is too large.
  For this reason, quantitativeness cannot be guaranteed in the case of $X$=N and C,
  and it is necessary to review the results using the first-principle calculations.
  However, qualitatively, it can be understood that the system exhibits a strong in-plane anisotropy.

  \subsection{MA Constants and Magnetization curves in SmFe$_{11}M$ ($M$=Ti, V, and Co)}

  \begin{figure}[htb]
    \begin{center}
      \includegraphics[width=7.0cm]{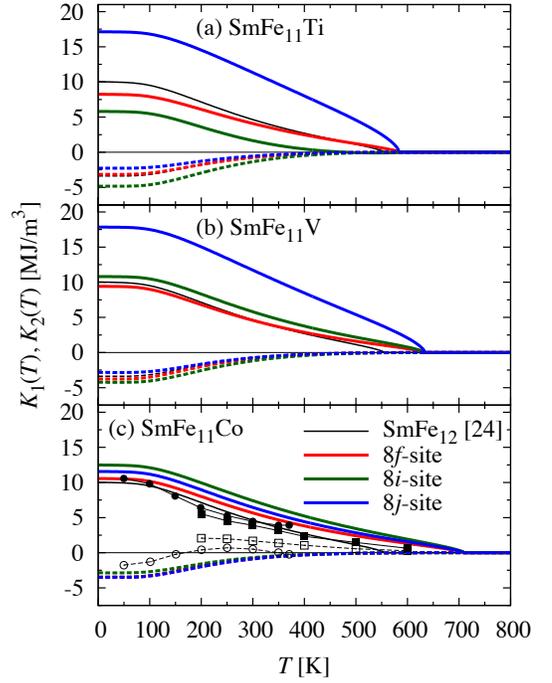}
      \caption{\label{fig:K1K2M}
        Temperature-dependent MA constants $K_1(T)$ (solid lines) and $K_2(T)$ (broken lines)
        [MJ/m$^3$] in SmFe$_{12}$ and SmFe$_{11}M$ [$M$=(a) Ti, (b) V, and (c) Co] compounds for each replacement site.
        These results are obtained using the analytical method for Eqs (21) and (22), mentioned in the  Ref. \cite{Yoshioka_SmFe12}.
        The experimental results for $K_1(T)$ and $K_2(T)$ in Sm(Co$_x$Fe$_{1-x}$)$_{12}$ are represented
        by solid and open plots, respectively, where
        the circles and squares represent the results for SmFe$_{11.16}$Co$_{0.84}$ using the Sucksmith-Thompson measurements \cite{Hirayama2}
        and for SmFe$_{10.8}$Co$_{1.2}$ measured using the anomalous Hall effect \cite{Ogawa}.}
    \end{center}
  \end{figure}

  The calculated MA constants $K_1$ and $K_2$ at finite temperatures are shown in FIG. \ref{fig:K1K2M}.
  At all temperatures, $K_1$ is positive and $K_2$ is negative.
  In case of substitution at the $8f$-site, which is the farthest from Sm,
  the influence of the $4f$ electron cloud is negligible.
  As a result, the temperature dependence is almost the same as that of the curve for SmFe$_{12}$,
  regardless of the type of the substituting element.
  This is consistent with the fact that a series of CF parameters and exchange fields have similar values.
  Conversely, when Fe($8j$) is replaced by Ti or V,
  $A_2^0\langle r^2\rangle$ takes a much larger value, and the rate of decrease of $K_1$ with an increasing temperature is small.
  This results from the fact that the generalized Brillouin function $B_J^l(x)$
  and the $T_J^l(x)$ function decrease more slowly with a smaller $l$ \cite{Yoshioka_SmFe12}.
  On the contrary, when Fe($8i$) is replaced by Ti,
  the rate of decrease of $K_1$ with an increasing temperature increases owing to the small value of $A_2^0\langle r^2\rangle$.
  Since, $K_2$ does not include the contribution of $A_2^0\langle r^2\rangle$, as shown in Eq. (\ref{eq:K2an}), the $K_2$ exhibits a negative temperature dependence in all cases and decays faster than $K_1$.

\begin{figure}[htb]
    \begin{center}
      \includegraphics[width=8.5cm]{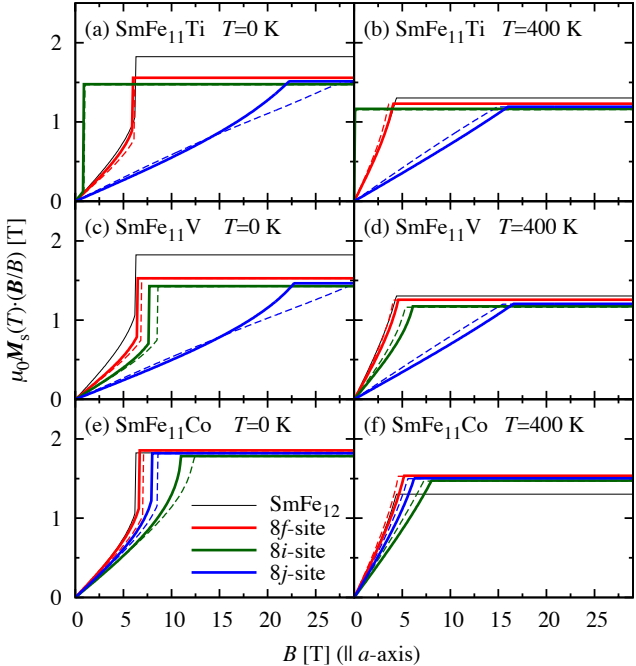}
      \caption{\label{fig:magcurveM}
        Magnetization curves along the $a$-axis at $T=0$ K and 400 K in SmFe$_{11}M$ [$M$=(a-b) Ti, (c-d) V, and (e-f) Co]
        compounds for each replacement site.
        These results are obtained by using the analytical method in Eq. (63) mentioned in the Ref. \cite{Yoshioka_SmFe12}.
        The thin broken curves show the statistical results obtained from Eq. (25) given in Ref. \cite{Yoshioka_SmFe12}.}
    \end{center}
  \end{figure}

  Similar to FIG. \ref{fig:magcurveX}, FIG. \ref{fig:magcurveM} shows the magnetization curve in the case of SmFe$_{11}M$ ($M$=Ti, V, and Co).
  Here we present the results for the case where the magnetic field is applied along the $a$-axis.
  Subsequently, there is a jump in the magnetization process for compounds that satisfy the FOMP condition at low temperatures.
  When Fe($8f$) is replaced by a transition-metal element, the magnetization curve is similar to that of SmFe$_{12}$.
  The values of the FOMP field ($T$=0 K) and the anisotropic field ($T$=400 K) are also close to those of SmFe$_{12}$.
  This is a consequence of the $8f$ site being the farthest Fe site from the Sm site as shown in FIG. \ref{fig:tlm}.
  However, the strong attraction of the rugbyball-like $4f$ electron cloud
  to the screened nuclear charge, as seen in the electronic structure analysis in Sec. \ref{sec:originM}, results in a significantly enhanced MA when Fe($8j$) is substituted with Ti or V.
  The large MA field of $H_{\rm A}\sim 16$ T is observed even at $T$=400 K
  because the decrease in the MA with an increasing temperature is slowed down by the large value of $A_2^{0}\langle r^2\rangle$.
  Although, when Fe($8i$) is replaced by Ti,
  FOMP is observed at low temperatures and small applied fields,
  and changes to an in-plane anisotropy at $T=400$ K with a zero field.
  This is consistent with the results shown in FIG. \ref{fig:K1K2M}
  from $K_{1,2}$(0)=5.80, -4.83 MJ/m$^3$ to $K_{1,2}(400)$=0.45, -0.47 MJ/m$^3$.
  In the case of Co-substitution, both the FOMP and MA fields exceed those of SmFe$_{12}$, regardless of the replacement site.
  Moreover, due to the large $T_{\rm c}$,
  the saturation magnetization of SmFe$_{11}$Ti exceeds that of SmFe12 at $T$=400 K.
  The magnetization curves i.e., the dashed curves, produced using the statistical method, are in good agreement with the present results,
  and thereby, confirm that the analysis of SmFe$_{11}$M is
  plausible within the framework of the linear theory of CF.
  The additional CF parameters have also been accounted for in the statistical calculations, as shown in TABLE \ref{tble:CFPres}.
  The details are presented in Appendix \ref{Sec:Aprm}.
  The results show that the effect of the breaking of local symmetry on the macroscopic magnetization curve is small under the action of strong exchange fields.

  \section{Summary}
  In this paper, the effect of element doping and substitution
  on the bulk magnetic properties of SmFe$_{12}$ compounds
  was studied by analyzing the electronic structure.
  The crystal fields, exchange fields, and magnetic moments were
  determined from the first-principle calculations, and consequently, an effective spin model was developed.
  The macroscopic magnetic properties of these SmFe$_{12}$ compounds were investigated using this model.
  The crystal field of the $4f$ electrons, generated due to the valence electron cloud
  and the screened nuclear charge, was investigated in detail,
  and the mechanism underlying the enhancement of the magnetic properties was clarified.
  We found that the first-order MA constant
  $K_1$ increases approximately by a factor of two
  when hydrogen is added to the $2b$ site and when Fe($8j$) is replaced by Ti or V.
  Moreover, we found that the decay of $K_1$ with an increasing temperature
  is slower than that in the other cases because $A_2^0$ has a particularly large value.
  This increase in the MA is realized
  by the attraction of the rugby-ball like $4f$ electron cloud
  to the shielded positive charge of the nucleus.
  The temperature dependence of $K_1(T)$ and $K_2(T)$
  in SmFe$_{11}$Co was found
  to qualitatively reproduce the experimental results for Sm(Fe$_{1-x}$Co$_x$)$_{12}$ ($x$=0.1, 0.07).
  Finally, the macroscopic magnetization curves were obtained
  from the electronic states within the first-order of the crystal field.
  Consequently, we found that the first-order magnetization process
  often appears in many SmFe$_{12}$-based compounds with a uniaxial MA
  that satisfy the condition $-K_2<K_1<-6K_2$ at low temperatures.
  Indeed,
  the compounds SmFe$_{12}$, SmFe$_{12}$H, SmFe$_{11}$Ti$_{8f}$, SmFe$_{11}$Ti$_{8i}$
  SmFe$_{11}$V$_{8f}$, SmFe$_{11}$V$_{8i}$, SmFe$_{11}$Co$_{8f}$, and SmFe$_{11}$Co$_{8j}$
  depicted the first-order magnetization process in our calculations, where the replacement site is indicated by
  the subscript.
  We confirmed that this scheme works well except for the case of SmFe$_{12}$C and SmFe$_{12}$N.

  \begin{acknowledgments}
    This work was supported by ESICMM Grant Number 12016013 and ESICMM is funded by
    Ministry of Education, Culture, Sports, Science and Technology (MEXT).
    T. Y. was supported by JSPS KAKENHI Grant Numbers JP21K04625.
    P. N. was supported by the project Solid21.
    Part of the numerical computations were carried out at the Cyberscience Center, Tohoku University, Japan.
  \end{acknowledgments}

  \appendix
  
  \section{Hamiltonian of a single $R$ Ion with the $LS$ Coupling Scheme}\label{Sec:HR}
 Here, we apply the $LS$ coupling scheme to
  the single $R$ Hamiltonian in Eq. (\ref{eq:Heff})
  owing to the strong Coulomb interaction between the $4f$ electrons.
  According to the Hund's rule for trivalent $R$ ion,
  we specify the quantum number of total orbital and spin moment $L$ and $S$
  for operators
  $\sum_{i=1}^{n_{4f}}\hat{\bm s}_i=\hat{\bm S}$ and $\sum_{i=1}^{n_{4f}}\hat{\bm l}_i=\hat{\bm L}$
  , respectively.
  The total angular momentum $J$ is varied from $|L-S|$ to $L+S$,
  and $M$ is the magnetic quantum number.
  Thus the single ion Hamiltonian in Eq. (\ref{eq:Heff})
  can be given as \cite{Yamada,Richter1}:
  \begin{align}
    \hat{\mathcal{H}}_{R}  =&
    \lambda\hat{\bm S}\cdot
    \hat{\bm L}
    +2\mu_{\rm B}\hat{\bm S}\cdot{\bm B}_{\rm ex}(T)
    +\sum_{l,m}\frac{A_{l}^{m}\langle r^l\rangle}{a_{l,m}}\Theta_{l}^{L} t_l^{m}(\hat{\bm L})\nonumber\\
    &+\mu_{\rm B}(\hat{\bm L}+\hat{\bm S})\cdot{\bm B},
    \label{HCF1}
  \end{align}
  where each term corresponds to
  $\hat{\cal H}_{{\rm so}}$, $\hat{\cal H}_{\rm ex}$, $\hat{\cal H}_{\rm CF}$, and
  $\hat{\cal H}_{\rm Z}$, respectively.
  Their corresponding basis can be written in the Russell-Saunders states $|L,S;J,M\rangle$.
  As for the spin-orbit interaction $\lambda$ in Sm ions,
  we use an experimental value of $\lambda/k_{\rm B}=411$ K \cite{Elliott}.
  In the $\hat{\cal H}_{\rm CF}$ term we use the following equivalent relation:\cite{Edomonds,Yoshioka_SmFe12}
  \begin{align}
    \sum_{i=1}^{n_{4f}} t_{l}^{m}(\hat{\theta}_{i},\hat{\phi}_{i})
    &=\Theta_{l}^{L} t_l^{m}(\hat{\bm L}),
  \end{align}
  with the factor:
  \begin{align}
    \Theta_{l}^{L}  &  =2^l\sqrt{\frac{(2L-l)!}{(2L+l+1)!}}
    \langle L\parallel\sum_{i=1}^{n_{4f}}C^{(l)}%
    (\hat{\theta}_{i},\hat{\phi}_{i})\parallel L\rangle,\label{eq:eqfct}
  \end{align}
  and operators:
  \begin{align}
    t_l^{\pm|m|}(\hat{\bm L})&=\sqrt{\pm\frac{2l+1}{8\pi}}
    \left[C_{-|m|}^{(l)}(\hat{\bm L})\pm(-1)^mC_{|m|}^{(l)}(\hat{\bm L})\right],\\
    t_l^{0}(\hat{\bm L})&=\sqrt{\frac{2l+1}{4\pi}}C_{0}^{(l)}(\hat{\bm L}),
  \end{align}
  for $m\ne 0$ and $m=0$, respectively.
  In the treatment of $\hat{\cal H}_{\rm so}$,
  we should note that
  because the $LS$ coupling in Sm compounds is weak compared with the other $R$ ones,
  the excited $J$-multiplets must be included\cite{Yoshioka_SmFe12,VanVleck,Sankar,Wijn,Kuzmin_mix,Magnani}.

  \section{Comparison of Total Energy in SmFe$_{11}M$ ($M$=Ti, V, and Co)}\label{Sec:energy}
In this section, we present the results in TABLE \ref{tble:toteng} showing the total energy per two f.u.
The most stable substitution sites were found to be $8i$, $8i$, and $8f$ sites for Ti, V, and Co substitutions, respectively.
In the case of Co substitution, the energy difference between $8f$ and $8j$ site substitutions was found to be small,
approximately 5.097 mRy/2f.u.
These results are consisent with the previous study \cite{Harashima3}
  \begin{table*}[htb]
    \caption{\label{tble:toteng}
      Comparison of the total energy in SmFe$_{11}M$ ($M$=Ti, V, and Co) for each replacement site,
      which is indicated by the subscript to $M$.}
    \begin{ruledtabular}
      \begin{tabular}{cccccc}
      $M$&Total Energy [Ry]&      $M$&Total Energy [Ry]&      $M$&Total Energy [Ry]
        \\ \hline
        Ti$_{8f}$& -101157.15591928 &
        V$_{8f}$ & -101539.19642501 &
        Co$_{8f}$& -103315.78997575\\
        Ti$_{8i}$& -101157.27273233 &
        V$_{8i}$ & -101539.25199749 &
        Co$_{8i}$& -103315.77202788\\
        Ti$_{8j}$& -101157.19639031 &
        V$_{8j}$ & -101539.20114312 &
        Co$_{8j}$& -103315.78487875\\
      \end{tabular}
    \end{ruledtabular}
  \end{table*}

  \section{Anisotropy Constants for Tetragonal Smmetry}\label{Sec:ki}

  According to the analytical method, using the modified effective lowest $J$-multiplet Hamiltonian,
  the MA energy for the $j$-th $R$ can be written explicitly in the form \cite{Yoshioka_SmFe12}:
  \begin{align}
    f_{{\rm CF},j}({\bm M}_{\rm s},T)=&\sum_{l,m}A_{l,j}^{m}\langle r^{l}\rangle\Xi_{l}^{J}\frac{t_{l}%
      ^{m}(\Theta,\Phi)}{a_{l,m}}\nonumber\\
    &\left[  J^{l}B_{J}^{l}(x_j)+
      \frac{l(l+1)}{2l+1}T_J^{l}(x_j)\right],\label{eq:FAan1}
  \end{align}
  with $x_j=-J(g_J-1)\mu_BB_{{\rm ex},j}(T)/k_{\rm B}T$,
  where $B_{{\rm ex},j}$ denotes the exchange field acting on a $4f$ shell in the $j$-th Sm
  and $g_J$ is Land\'e $g$ factor.
  $B_J^l(x_j)$ is the generalized Brillouin function \cite{Kuzmin_linear,Magnani}
  and $T_J^l(x_j)$ is the function defined by Eq. (50) in Ref. \cite{Yoshioka_SmFe12}.
  The equivalent factor $\Xi_l^J$ in Eq. (\ref{eq:FAan1}) was introduced in Ref. \cite{Yoshioka_SmFe12},
  and can be written in the form:
  \begin{align}
    \Xi_l^J=&2^l\sqrt{\frac{(2J+1)(L+1)}{S}}\sqrt{\frac{(2J+l+2)(2J-l+1)!}{l(l+1)(2J+l+1)!}}\nonumber\\
    &\times
    \begin{Bmatrix}
      L   & J & S\\
      J+1 & L & l
    \end{Bmatrix}
    \langle L\parallel\sum_{i=1}^{n_{4f,j}}C^{(l)}%
    (\hat{\theta}_{i},\hat{\phi}_{i})\parallel L\rangle.
  \end{align}

  By comparing $f_{{\rm CF},j}({\bm M}_s,T)-f_{{\rm CF},j}(M_s{\bm n}_c,T)$ in Eq. (\ref{eq:FAan1}) with first term on the right hand side of Eq. (\ref{eq:Fan}), the following
  MA constants with a four-fold rotational symmetry
  for trivalent magnetic light $R$ ion
  (Ce$^{3+}$, Pr$^{3+}$, Nd$^{3+}$, Pm$^{3+}$, and Sm$^{3+}$) can be written as follows:
  \begin{eqnarray}
    k_{1}(T) &=&-3\left[  J^{2}B_{J}^{2}(x)+\frac{6}{5}%
      T_J^{2}(x)\right]  A_{2}^{0}\langle
    r^{2}\rangle\Xi_{2}^{J}\nonumber\\
    &&  -40\left[  J^{4}B_{J}^{4}(x)+\frac{20}{9}%
      T_J^{4}(x)\right]  A_{4}^{0}\langle
    r^{4}\rangle\Xi_{4}^{J}\nonumber\\
    &&  -168\left[  J^{6}B_{J}^{6}(x)+\frac{42}{13}%
      T_J^{6}(x)\right]  A_{6}^{0}\langle
    r^{6}\rangle\Xi_{6}^{J},\label{eq:K1an}\\
    k_{2}(T) &  =&35\left[  J^{4}B_{J}^{4}(x)+\frac{20}{9}
      T_J^{4}(x)\right]  A_{4}%
    ^{0}\langle r^{4}\rangle\Xi_{4}^{J}\label{eq:K2an}\nonumber\\
    &&  +378\left[  J^{6}B_{J}^{6}(x)+\frac{42}{13}%
      T_J^{6}(x)\right]  A_{6}^{0}\langle
    r^{6}\rangle\Xi_{6}^{J},\\
    k_{2}^1(T) &  =&\left[  J^{4}B_{J}^{4}(x)+\frac{20}{9}
      T_J^{4}(x)\right]  A_{4}%
    ^{4}\langle r^{4}\rangle\Xi_{4}^{J}\nonumber\\
    &&  +10\left[  J^{6}B_{J}^{6}(x)+\frac{42}{13}%
      T_J^{6}(x)\right]  A_{6}^{4}\langle
    r^{6}\rangle\Xi_{6}^{J}, \label{eq:k21}\\
    k_{3}(T) &  =&-231\left[  J^{6}B_{J}^{6}(x)+\frac{42}{13}%
      T_J^{6}(x)\right]  A_{6}^{0}\langle
    r^{6}\rangle\Xi_{6}^{J},\\
    k_{3}^1(T) &  =&-11\left[  J^{6}B_{J}^{6}(x)+\frac{42}{13}%
      T_J^{6}(x)\right]  A_{6}^{4}\langle
    r^{6}\rangle\Xi_{6}^{J}.
  \end{eqnarray}

  \section{Symmetry Operations on SmFe$_{11}M$ ($M$=Ti, V, and Co)}\label{Sec:Aprm}

  \begin{table*}[t]
    \caption{\label{tble:CFPres}
      Crystal field parameters $A_{l,1}^m\langle r^l\rangle$ [K] for ($l,m$)
      without a four-fold rotational symmetry
      in SmFe11$M$ ($M$=Ti, V, and Co)
      for each replacement site, which is indicated by the subscript to $M$.}
    \begin{ruledtabular}
      \begin{tabular}{cccccccccccccccccccc}
        system&
        (2, $\pm$ 1)&
        (2,-2)&
        (2, 2)&
        (4, $\pm$ 1)&
        (4,-2)&
        (4, 2)&
        (4, $\pm$ 3)&
        (6, $\pm$ 1)&
        (6,-2)&
        (6, 2)&
        (6, $\pm$ 3)&
        (6, $\pm$ 5)&
        (6,-6)&
        (6, 6)
        \\ \hline
        Ti$_{8f}$& $\mp$25.5 & 88.9 && $\pm$57.0 & -24.7 && 110.8 &$\mp$5.4 &-5.4 && -11.1 & $\pm$41.9 & -13.4 \\
        Ti$_{8i}$&&& -320.8 &&& 52.7  &&&& -5.3 &&&& -10.4 &\\
        Ti$_{8j}$&&& 164.2  &&& 138.8 &&&& 9.5  &&&& 0.8   &\\

        V$_{8f}$ & $\mp$35.2 &80.2  && $\pm$24.4 & -18.6 && 57.1  &$\mp$3.8 &-3.3 && -7.7  & $\pm$27.1 & -8.7  \\
        V$_{8i}$ &&& -285.1 &&& 28.5 &&&& -3.7 &&&& -8.8 & \\
        V$_{8j}$ &&& 146.5  &&& 74.7 &&&& 6.9  &&&& 1.4  &\\

        Co$_{8f}$& $\mp$124.6& 80.7 && $\mp$7.0  & -3.5 && 13.8   &$\pm$1.4 & 1.4 && 2.2   & $\mp$12.5 & 3.4   \\
        Co$_{8i}$&&& -42.1  &&& -21.3 &&&& 0.8  &&&& 1.7   & \\
        Co$_{8j}$&&& 36.0   &&& -29.5 &&&& -2.8 &&&& 1.0   &\\
      \end{tabular}
    \end{ruledtabular}
  \end{table*}

  Elemental substitutions locally break the four-fold rotational symmetry,
  which is considered to be recovered in the whole system.
  In this section, we describe the procedure to calculate
  the bulk magnetic properties from the CF parameters obtained using the first-principle calculations
  for the specific structure shown in FIG. \ref{fig:struct} (b-d),
  in which the complete set of CF parameters include the contribution shown in TABLE \ref{tble:CFPres}.
  The Hamiltonian for a system satisfying a four-fold rotational symmetry can be written in general as:
  \begin{align}
    \hat{\cal H}=&\frac{1}{4}\sum_{k=1}^{4}\sum_{j=1}^{n_R}e^{i\hat{L}_z\pi (k-1)/2}
    \hat{\cal H}_{R,j} e^{-i\hat{L}_z\pi (k-1)/2}\nonumber\\
    &+VK_{1}^{\rm TM}(T)\sin^{2}%
    \theta^{\rm TM}-V\boldsymbol{M}^{\rm TM}(T)\cdot\boldsymbol{B}.
  \end{align}
  In particular, by applying rotational operations to the CF Hamiltonian,
  the CF parameters at the $k$-th $R$ sites can be obtained in the following form:
  \begin{align}
    e^{i\hat{L}_z\pi (k-1)/2}
    \hat{\cal H}_{{\rm CF}} e^{-i\hat{L}_z\pi (k-1)/2}
    &=\sum_{l,m}\frac{A_{l,k}^{m}\langle r^l\rangle}{a_{l,m}}\Theta_{l}^{L} t_l^{m}(\hat{\bm L}_k)
  \end{align}
  with
  \begin{align}
    A_{l,k}^0\langle r^l\rangle     =&A_l^0\langle r^l\rangle\\
    A_{l,k}^{-|m|}\langle r^l\rangle=
    & \cos\left[\frac{\pi}{2}|m|(k-1)\right]A_{l}^{-|m|}\langle r^l\rangle \nonumber\\
    &-\sin\left[\frac{\pi}{2}|m|(k-1)\right]A_{l}^{|m|}\langle r^l\rangle\\
    A_{l,k}^{|m|}\langle r^l\rangle =
    & \sin\left[\frac{\pi}{2}|m|(k-1)\right]A_{l}^{-|m|}\langle r^l\rangle\nonumber\\
    &+\cos\left[\frac{\pi}{2}|m|(k-1)\right]A_{l}^{|m|}\langle r^l\rangle.
  \end{align}

  In the case of the analytical calculations using linear theory,
  the total MA energy can be written
  as a linear combination of $A_{l,k}^m$, and the expected value of
  the spherical tensor operator $\langle C^{(l)}_0(\hat{\bm {L}}_k)\rangle$ is common
  for each $k$-th $R$ ion.
  Thus, all the terms including $A_{l,k}^m$, shown in TABLE \ref{tble:CFPres},
  cancel and do not contribute to the macroscopic MA.
  Conversely,
  each $R$ exhibits a non-collinear structure independently relative
  to the partial magnetization ${\bm M}^{\rm TM}(T)$ in the numerical diagonalization and statistical calculations.
  Therefore, the complete cancellation of the $A_{l,k}^m$ term in TABLE \ref{tble:CFPres},
  which can be seen in the analytical formula, does not occur.

\begin{figure}[H]
    \begin{center}
      \includegraphics[width=7.0cm]{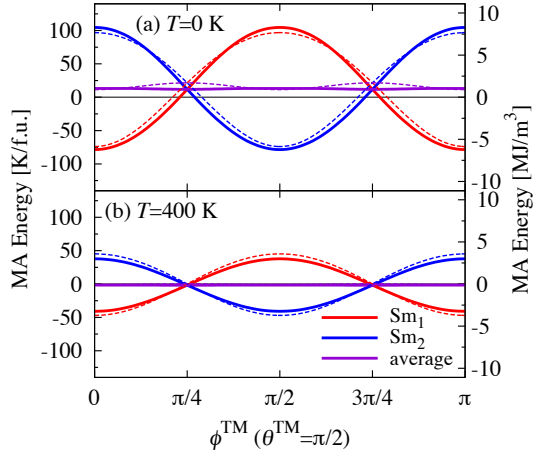}
      \caption{\label{fig:energy_angle}
        Angular dependence of MA energy at (a) $T=0$ K and (b) $T=400$ K
        for two inequivalent sites and their averaged value (Sm$_{1,2}$ and average)
        in SmFe$_{11}$Ti with 8$i$ substitution.
        $\theta^{\rm TM}$ and $\phi^{\rm TM}$ are the polar and azimuthal angle of partial magnetization ${\bm M}^{\rm TM}$.
        The origin of the energy is taken at $\theta^{\rm TM}=0$.
        These results are obtained by applying Eq. (14) at ${\bm B}=0$ of Ref. \cite{Yoshioka_SmFe12}.}
    \end{center}
  \end{figure}

  For SmFe$_{11}$Ti with the $8i$ substitution, which is the most affected by the Ti substitution,
  the MA energy calculated using the numerical diagonalization and analytical formulae are represented
  by solid and dashed curves in FIG. \ref{fig:energy_angle}, respectively.
  Due to the large value of $A_{2,1(2)}^2\langle r^2\rangle$=-320.8 (320.8) K,
  Sm$_{1(2)}$ exhibits a strong in-plane anisotropy proportional to
  the form $A_{2,1(2)}^2\cos 2\phi^{\rm TM}$.
  Therefore, these contributions completely cancel each other out.
  The result of numerical diagonalization also cancels the contribution to the same extent
  as in the analytical calculations.
  As a result, we can confirm that the MA energy of the bulk is well approximated
  by the analytical calculation method, where a collinear structure between the partial magnetization
  and the magnetic moment of each Sm ion is assumed.

  \end{document}